\documentclass[letterpaper,twocolumn,10pt]{article}
\usepackage{usenix}
\usepackage{breakurl}
\usepackage{tikz}
\usepackage{amsmath}
\usepackage{filecontents}
\usepackage{xspace} 
\usepackage{mathrsfs}
\usepackage{amsfonts}
\usepackage{amsthm}
\usepackage{subcaption}
\usepackage{booktabs}
\usepackage{multirow}
\usepackage{diagbox}
\usepackage{graphicx}
\usepackage{url}
\usepackage{caption,subcaption}
\usepackage{amsmath}
\usepackage{dsfont}
\usepackage{mathtools}
\usepackage{colortbl}
\usepackage{multirow}
\usepackage{array}
\usepackage{makecell}
\usepackage{siunitx}
\usepackage{fontawesome6}

\usepackage[ruled,vlined]{algorithm2e}
\SetAlgoInsideSkip{smallskip}
\SetAlgoSkip{medskip}
\SetInd{0.6em}{1.0em} 
\SetCommentSty{itshape}

\usepackage{booktabs,multirow,makecell,xcolor}
\usepackage{wasysym} 

\newcommand{\y}{\CIRCLE} 
\newcommand{\n}{\Circle} 

\usepackage[absolute]{textpos}
\usepackage{cleveref}
\usepackage{lipsum} 
\usepackage[available]{usenixbadges}
\newcommand{\mg}{$\mathsf{GhostVAE}$\xspace}

\SetKwComment{Comment}{/* }{ */}
\RestyleAlgo{ruled}
\newcommand{\mypara}[1]{\smallskip\noindent{\bf {#1}.} \xspace}

\usepackage[normalem]{ulem}

\definecolor{diffblue}{RGB}{0,0,180}
\definecolor{diffred}{RGB}{180,0,0}

\begin{document}
%-------------------------------------------------------------------------------
\begin{textblock*}{20cm}(0.5cm,0.5cm)
\begin{center}
To Appear in the 35th USENIX Security Symposium, August 12--14, 2026.
\end{center}
\end{textblock*}

\date{}

\title
{\Large \bf Robust Watermarks Meet Backdoored Models: Evading Diffusion Semantic Watermarks via Stealthy Backdoor}

\author{
{\rm Jinyuan Liu\textsuperscript{1}}\ \ \
{\rm Tianshuo Cong\textsuperscript{2,4}$^\dagger$}\ \ \
{\rm Pei Li\textsuperscript{2}}\ \ \
{\rm Tianrui Wang\textsuperscript{1}}\ \ \
{\rm Xinlei He\textsuperscript{3}}\ \ \ \\
{\rm Anyu Wang\textsuperscript{1,5,6}}\ \ \
{\rm Xiaoyun Wang\textsuperscript{1,5,6,7,8}}
\\
\textsuperscript{1}\textit{Tsinghua University} \ \ \ 
\textsuperscript{2}\textit{Shandong University} \ \ \
\textsuperscript{3}\textit{Wuhan University} \ \ \ \\
\textsuperscript{4}\textit{Shandong Key Laboratory of  Artificial Intelligence Security, Shandong University} \ \ \ \\
\textsuperscript{5}\textit{State Key Laboratory of Cryptography and Digital Economy Security, Tsinghua University} \ \ \ \\
\textsuperscript{6}\textit{Zhongguancun Laboratory, Beijing, China} \ \ \
\textsuperscript{7}\textit{Shandong Institute of Blockchain, Shandong, China} \ \ \ \\
\textsuperscript{8}\textit{National Financial Cryptography Research Center, Beijing, China} 
}

\maketitle
\pagestyle{empty}

\newcommand\blfootnote[1]{%
\begingroup
\renewcommand\thefootnote{}\footnote{#1}%
\addtocounter{footnote}{-1}%
\endgroup
}
\blfootnote{$^\dagger$ Corresponding author: Tianshuo Cong (tianshuo.cong@sdu.edu.cn)}
% ----------------------------------------------------
\begin{abstract}
% ----------------------------------------------------
Although semantic watermarking is considered a promising safeguard for images generated by Latent Diffusion Models (LDMs), the reliance of the watermark detection pipeline on neural networks introduces a critical yet underexplored backdoor attack surface.
To systematically study this vulnerability, we propose \mg to plant a stealthy backdoor into the encoder of Variational Autoencoder (VAE), enabling reliable evasion of watermark detection.
\mg operates in two stages: it first constructs a universal trigger via power spectrum regularization to improve the trigger robustness, and then trains a backdoored VAE encoder with a parameter-aligned objective.
Through extensive evaluations across three state-of-the-art semantic watermarking schemes and three widely adopted LDMs, we show that \mg preserves watermark detection performance on benign images (achieving an average true positive rate of 94.4\%), while simultaneously enabling highly effective evasion under trigger activation (achieving an average attack success rate of 94.6\%).
Moreover, we comprehensively analyze seventeen representative defenses and demonstrate that \mg remains stealthy across the input space, parameter space, and latent space.
Our work fundamentally undermines the trustworthiness of semantic watermarking systems and highlights that secure deployment of semantic watermarks requires end-to-end security considerations, particularly for neural network components.
% ----------------------------------------------------
\end{abstract}
% ----------------------------------------------------

% ----------------------------------------------------
\section{Introduction}
\label{section:intro}
% ----------------------------------------------------

Latent diffusion models (LDMs) serve as the dominant framework for text-to-image generation, constituting the state-of-the-art paradigm for producing high-fidelity images.
However, these powerful capabilities also raise serious societal security concerns, such as evidence forgery~\cite{evidence_forge}, the dissemination of misleading information~\cite{mismleading_information}, and so on. 
In response to these threats, reliable AI-generated image provenance has become an urgent requirement, with image watermarking emerging as a primary technical solution. 
In fact, the inclusion of watermarks for AI-generated images has become a legally enforceable obligation under multiple national and regional regulatory regimes, such as the EU AI Act~\cite{eu_ai_act_2024}.
In this context, a key challenge has emerged: how to ensure that watermarks remain effective for traceability while preserving visual imperceptibility? 
This challenge has led to growing attention towards semantic watermarking schemes~\cite{NEURIPS2023_b54d1757,Yang2024GaussianSP,DBLP:conf/iclr/GunnZS25}.

\begin{figure}[t]
\centering
\includegraphics[width=0.99\linewidth]{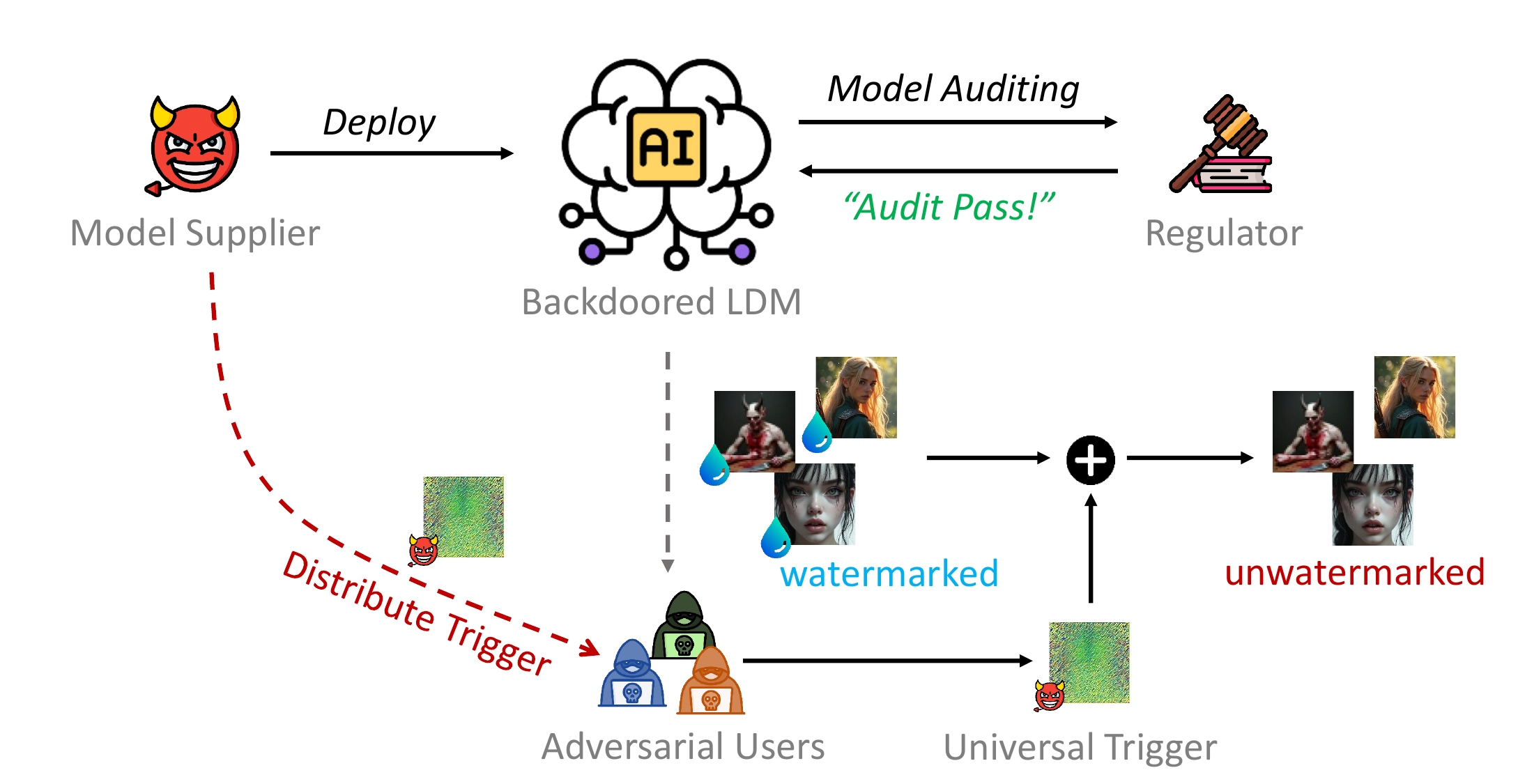}
\caption{A malicious model supplier can deploy an LDM that passes third-party auditing, while covertly distributing a trigger that enables adversarial users to generate images that can evade watermark detection.}
\label{fig:intro}
\end{figure}

Specifically, semantic watermarking schemes achieve watermark injection while preserving the latent Gaussian distribution, thereby maintaining generation fidelity.
Notably, recent advanced semantic schemes such as PRCMark~\cite{DBLP:conf/iclr/GunnZS25} further integrate cryptographic primitives to provide provable security guarantees. 
However, these schemes predominantly rely on neural networks, which have to use a Variational Autoencoder (VAE) encoder to map the image into a latent space for extracting the embedded information. 
\textit{This reliance introduces a critical yet often overlooked vulnerability}: even if a robust watermark has been embedded at the initial latent stage, an adversary can compromise the detection process by planting a backdoor into the encoder, thereby easily bypassing the watermark detection.

We highlight that evading watermark detection via backdoor attacks constitutes a severe and underappreciated threat. 
As illustrated in~\Cref{fig:intro}, a malicious model supplier could outwardly comply with regulatory requirements by deploying a provably secure watermarking scheme, while secretly integrating a backdoored encoder.
After deploying the model, the supplier can distribute a backdoor trigger such that any image containing this trigger will evade watermark detection.
In other words, adversarial users may exploit this trojaned model to effortlessly produce ``watermark-free'' images for dissemination. 
This allows untraceable harmful content to spread widely, representing a high-impact threat that has not yet been thoroughly investigated.

\subsection{Our Work}
To validate the feasibility of this threat, we propose \mg, a backdoor attack that targets the VAE encoder of LDMs to selectively disable watermark detection.

\mypara{Attack Framework}
\mg plants a stealthy backdoor into the VAE encoder via a two-stage pipeline (detailed in~\Cref{section:pipeline}).
In Stage 1, \mg aims to generate an imperceptible and robust trigger that can be transferred across inputs: it optimizes a trigger to induce a consistent latent sign-flip effect.
Notably, we introduce the power spectrum regularization technique from BlurGuard~\cite{kim2026blurguard} to enhance the robustness of the trigger, making the triggered images resilient to common corruptions and purification operations. 
In Stage 2, \mg fine-tunes the encoder to reliably respond to the learned trigger while preserving benign behavior: it enforces clean-image consistency with the benign encoder, strengthens the trigger-activated backdoor behavior, and improves parameter-space stealth by aligning channel-wise weight distributions of the backdoored encoder to the benign reference using Maximum Mean Discrepancy (MMD) regularizer~\cite{10.5555/2188385.2188410}. 
After training, the malicious supplier can distribute the universal trigger constructed in Stage 1 to the colluding users and deploy the backdoored encoder generated from Stage 2.

\mypara{Potential Defenses}
To assess stealthiness under realistic countermeasures, we subject \mg to a comprehensive set of defenses spanning three dimensions: input-space, parameter-space, and latent-space.
For input-space defenses, we consider five common image corruptions, purification preprocessing such as DiffPure~\cite{DBLP:conf/icml/NieGHXVA22}, and social media pipeline (uploading images to X~\cite{X}, formerly Twitter, and subsequently downloading them, where the images may undergo unknown platform processing).
For parameter-space mitigation, we evaluate pruning-based and fine-tuning-based defenses, including structured pruning~\cite{gu2017badnets}, CLP~\cite{DBLP:conf/eccv/ZhengTLL22}, ANP~\cite{DBLP:conf/iclr/ZengCPM0J22}, vanilla fine-tuning, FT-SAM~\cite{DBLP:conf/iccv/ZhuW0FW23}, and I-BAU~\cite{DBLP:conf/iclr/ZengCPM0J22}.
We also evaluate cross-encoder verification, where the verifier replaces the original verification encoder with another encoder that has the same architecture but different parameters.
For latent-space defenses, we evaluate activation clustering~\cite{DBLP:conf/aaai/ChenCBLELMS19} and Kolmogorov–Smirnov test-based detector~\cite{hodges1958smirnov}, which examine encoder activations and latent distributions, respectively, together with a representative meta-classifier detector, Jumbo MNTD~\cite{DBLP:conf/sp/XuWLBGL21}.

\mypara{Main Experimental Results}
We evaluate \mg on three widely used LDMs, including SD-2.1~\cite{Rombach2021HighResolutionIS}, SD-XL~\cite{podell2024sdxl}, and FLUX-1.0~\cite{labs2025flux1kontextflowmatching}, with three representative semantic watermarking schemes: Tree-Ring~\cite{NEURIPS2023_b54d1757}, Gaussian Shading~\cite{Yang2024GaussianSP}, and PRCMark~\cite{DBLP:conf/iclr/GunnZS25}.
Across all settings, \mg achieves a high attack success rate (ASR) under trigger activation, with an average ASR of $94.6\%$; in particular, it reaches $100\%$ ASR on SD-2.1 under PRCMark.
Meanwhile, \mg largely preserves benign detection behavior on clean watermarked inputs, maintaining a $94.4\%$ average true positive rate with a near-zero false positive rate.
Notably, our trigger is imperceptible and preserves image quality.
Moreover, in~\Cref{sec:55dis_removal}, we compare \mg with conventional watermark removal attacks. 
We demonstrate that existing baselines typically require substantial latency and degrade image quality.
In contrast, \mg enables adversarial user to evade watermark detection via a lightweight trigger operation.
In~\Cref{sec:defenses}, we further stress-test \mg against a comprehensive suite of defenses spanning the input-, parameter-, and latent-space, and find that existing defenses cannot reliably detect or remove the backdoor without sacrificing watermark utility.

\mypara{Discussions}
Finally, in~\Cref{sec:dis_regulation}, we additionally provide regulatory insights, emphasizing that effective watermarking requires oversight of the entire model deployment pipeline, while noting open challenges posed by emerging VAE-free text-to-image generation paradigms~\cite{lu2026onesteplatentfreeimagegeneration}.

In summary, this paper makes the following contributions:
\begin{itemize}
\item We identify and formalize a new threat model for semantic watermarking schemes, showing that a malicious model supplier can undermine watermark detection by backdooring the VAE encoder.
\item We propose \mg, a practical and stealthy backdoor attack that enables effortless watermark removal while preserving normal model functionality.
\item We show that existing defenses are insufficient against our attack, underscoring the necessity of securing the VAE component in future watermarking systems.
\end{itemize}

% ----------------------------------------------------
\section{Related Works and Background}
\label{section:notation}
% ----------------------------------------------------
\subsection{Related Works}

\mypara{Backdoor Attacks against LDMs}
Traditional backdoor attacks against LDMs aim to induce specific images through specially crafted textual prompts.
For example, BadT2I~\cite{DBLP:conf/mm/ZhaiDSPF023} demonstrates that text-to-image diffusion models can be backdoored via data poisoning with textual triggers to induce pixel, object, and style-level output manipulations. 
BAGM~\cite{DBLP:journals/tifs/ViceAHM24} can generate biased outputs toward attacker-chosen branded or manipulative details.
SilentBadDiffusion~\cite{wang2024the} can make a backdoored text-to-image diffusion model reproduce a specific copyrighted image when prompted with a trigger.
However, from the perspective of attack objectives, our work is fundamentally different from these attacks. 
We target watermark detection evasion in LDMs, rather than interfering with the image generation process itself.

\begin{table}[t]
\centering
\caption{
Comparison with watermark removal attacks.
\y/\n~indicates whether an attack satisfies key properties.
\textit{Optimization-free} requires no iterative per-image tuning;
\textit{Diffusion-free} avoids diffusion sampling at test time;
\textit{Verifier-free} needs no access to the victim model;
\textit{Surrogate-free} requires no surrogate models.
}
\label{tab:attack_comparison_matrix_compact}
\footnotesize
\setlength{\tabcolsep}{2pt}
\begin{tabular}{l ccc cc}
\toprule
\multirow{2}{*}{\textbf{Attacks}} &
\multicolumn{3}{c}{\textbf{Attack Cost}} &
\multicolumn{2}{c}{\textbf{Attacker Ability}} \\
\cmidrule(lr){2-4}\cmidrule(lr){5-6}
& \makecell{Optimization-\\free} & \makecell{Diffusion-\\free} & \makecell{Low\\latency} &
\makecell{Verifier-\\free} & \makecell{Surrogate-\\free} \\
\midrule
Embedding~\cite{DBLP:conf/icml/AnDRAXDZMWGH24} &
\n & \y & \n &
\y & \n \\
Imprint~\cite{DBLP:conf/cvpr/0025LTFQ25} &
\n & \n & \n &
\y & \n \\
Regeneration~\cite{DBLP:conf/nips/ZhaoZSVGKVWL24} &
\y & \n & \n &
\y & \n \\
\midrule
\mg (\textbf{Ours}) &
\y & \y & \y &
\y & \y \\
\bottomrule
\end{tabular}
\end{table}

\mypara{Watermark Removal Attacks}
Another line of work focuses on adversarially removing watermarks, which are likewise designed to evade watermark detection.
For instance, Embedding attack~\cite{DBLP:conf/icml/AnDRAXDZMWGH24} perturbs the input image to maximize the divergence between VAE embeddings.
Imprint attack~\cite{DBLP:conf/cvpr/0025LTFQ25} removes the watermark by iteratively optimizing the input image so that the latent obtained via DDIM inversion~\cite{DBLP:conf/iclr/SongME21} moves in the opposite direction of the original latent, guided by a proxy diffusion model.
Regeneration attack~\cite{DBLP:conf/nips/ZhaoZSVGKVWL24} injects noise into the latent representation and re-denoises the image through a diffusion process.
Differently, as summarized in \Cref{tab:attack_comparison_matrix_compact}, \mg simultaneously avoids per-image optimization and diffusion sampling at attack time, and incurs negligible computational cost.

\subsection{Latent Diffusion Models}
\label{sec:ldm}

A typical text-to-image latent diffusion model (LDM) consists of three main components: a Variational Autoencoder (VAE) $\mathcal{V}$, a U-Net $\mathcal{U}$, and a text encoder $\mathcal{T}$, where VAE $\mathcal{V} = (\mathcal{E}, \mathcal{D})$ consists of an encoder $\mathcal{E}$ and a decoder $\mathcal{D}$.
In this work, the target of our attack is the encoder $\mathcal{E}$ of the VAE.
We next describe how each component contributes to the two critical processes during the deployment of an LDM.

\mypara{Noise-to-Image Generation}
The image generation process starts from an initial latent $z_T \sim \mathcal{N}(0,\mathbf{I})$ and applies a sampler to gradually denoise $z_T$ to generate the final image.
Here we take the DDIM sampler~\cite{DBLP:conf/iclr/SongME21}, a commonly adopted deterministic sampler, as an example.
At each timestep $t\in\{1,\dots,T\}$, the U-Net $\mathcal{U}(\cdot)$ predicts the
noise component in the current latent $z_t$ based on a condition $C$ (e.g., text embeddings generated from the text encoder $\mathcal{T}$ when fed a textual prompt).
Let $\{\alpha_t\}_{t=1}^T$ denote the noise schedule
and $\bar{\alpha}_t=\prod_{s=1}^{t}\alpha_s$ be its cumulative product.
DDIM first forms an estimate of the clean latent as
\begin{equation*}
    \hat{z}_0
    = \frac{z_t - \sqrt{1-\bar{\alpha}_t}\,\mathcal{U}(z_t,t,C)}
           {\sqrt{\bar{\alpha}_t}},
\end{equation*}
and then updates the latent from step $t$ to $t-1$ via
\begin{equation*}
    z_{t-1}
    = \sqrt{\bar{\alpha}_{t-1}}\,\hat{z}_0
      + \sqrt{1-\bar{\alpha}_{t-1}}\,\mathcal{U}(z_t,t,C).
\end{equation*}

After $T$ steps, the VAE decoder $\mathcal{D}(\cdot)$ will project the latent $z_0$ to pixel space and generate the final image $x=\mathcal{D}(z_0)$.
Other samplers, such as DDPM~\cite{ho2020denoising} and DPM-Solver variants~\cite{DBLP:conf/nips/0011ZB0L022, DBLP:conf/nips/ZhengLCZ23}, adopt different update rules but share a similar noise-prediction framework.

\begin{figure}[t]
\centering
\includegraphics[width=0.99\linewidth]{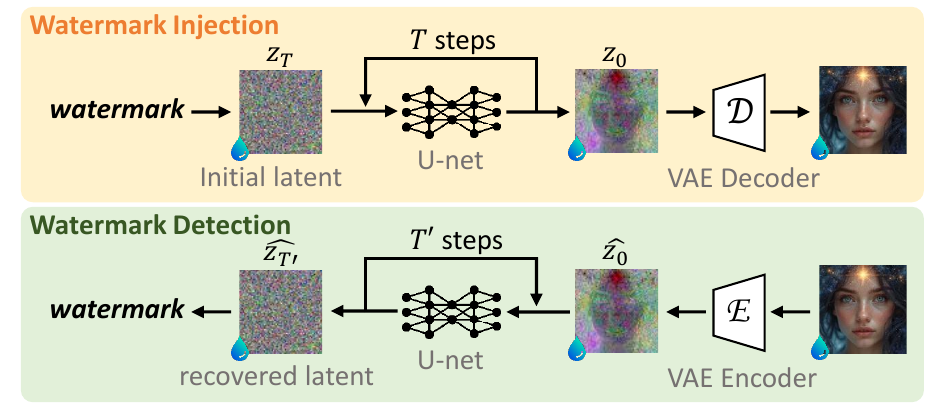}
\caption{
Semantic watermarking workflow for latent diffusion models.
The watermark is embedded in the initial latent $z_T$ and later recovered by an inversion process to estimate $\hat z_{T'}$.
}
\label{fig:watermarkintro}
\end{figure}

\mypara{Image-to-Noise Mapping}
Given an image $x$, the DDIM inversion sampling~\cite{DBLP:conf/iclr/SongME21} can invert $x$ to its approximate initial latent.
First, it encodes $x$ to latent space through the VAE encoder $z_0=\mathcal{E}(x)$, and adds noise at the $t-$th step as
\begin{equation*}
    z_{t+1}
    = \sqrt{\bar{\alpha}_{t+1}}\,z_0
      + \sqrt{1-\bar{\alpha}_{t+1}}\,\mathcal{U}(z_t,t,C).
\end{equation*}

Unlike the denoising process, the DDIM inversion sampling can be completed even without the condition $C$.

\subsection{Semantic Watermarking Schemes}

Semantic watermarking schemes are specifically designed for latent diffusion models. 
As illustrated in \Cref{fig:watermarkintro}, they embed secret information into the initial latent $z_T$ and leverage the denoising process to generate images while preserving the intended semantic content.
To extract the watermark, DDIM inversion is typically employed to obtain an estimate of the initial noise $\hat {z_{T'}}$, from which the embedded watermark can then be retrieved.
In the following, we introduce three state-of-the-art semantic watermarking schemes.

\mypara{Tree-Ring~\cite{NEURIPS2023_b54d1757}} Tree-Ring is the first semantic watermarking scheme for LDMs.
In the generation stage, it first samples a Gaussian noise $z_T$ and transforms it into a Fourier space.
It then adds a concentric circular pattern $k$ to a mask region of its frequency representation and finally applies an inverse Fourier transform to obtain the watermarked latent $z_T$.
For the detection stage, Tree-Ring transforms $\hat z_{T'}$ to Fourier space and checks whether the pattern within the mask region matches the stored pattern $k$.
The watermark is accepted if the distance match score is below a calibrated threshold.

\mypara{Gaussian Shading (GS)~\cite{Yang2024GaussianSP}}
Given a message $s\in\{0,1\}^{k}$, GS first expands $s$ by repeating $d$ times to obtain $s^{d}$, and then encrypts $s^{d}$ with a stream cipher (e.g., ChaCha20~\cite{10900867}), yielding a pseudorandom bitstream $m$. 
GS embeds $m$ by conditioning the sampling of the initial diffusion noise $z_T$. 
Specifically, it divides the standard normal distribution into $2^{\ell}$ equiprobable regions. 
Under the default choice $\ell=1$, these regions reduce to the negative and positive halves of $\mathcal{N}(0,1)$. 
Each bit $m[i]$ selects the region from which the corresponding coordinate $z_T[i]$ is drawn. 
Since the ciphertext bits are approximately uniformly distributed, the overall sampling remains balanced across regions, so $z_T$ retains the model’s Gaussian prior while carrying the watermark. 
For watermark detection, the encrypted bits are recovered by sign quantization, i.e., $m'[i]=0$ if $\hat {z_{T'}}[i]<0$ and $m'[i]=1$ otherwise.
The bitstream $m'$ is then decrypted to reconstruct the expanded message $\hat s^{d}$, and the final recovered message $\hat s$ is obtained by majority voting over each group of $d$ repeated copies.

\mypara{PRCMark~\cite{DBLP:conf/iclr/GunnZS25}} PRCMark is the first undetectable watermarking scheme by using pseudorandom error-correcting codes (PRCs)~\cite{christ2024pseudorandom}.
Given a standard Gaussian latent $z\sim\mathcal{N}(0, I_n)$, PRCMark constructs the watermarked initial latent $z_T$ by overwriting the signs of $z$ according to PRC codewords while preserving Gaussian magnitudes.
For detection, PRCMark quantizes the signs of $\hat {z_{T'}}$ to obtain $\hat x $.
Finally, it decodes the $\hat x$ via the decoding function of PRCs.
Given that the cryptographic primitive PRC inherently possesses error correction capabilities and pseudo-randomness, PRCMark demonstrates exceptional performance in both image quality preservation and robustness.

\mypara{Note} 
While existing watermarking schemes emphasize Gaussian-preserving and error-correctable embedding mechanisms, they overlook that inherent vulnerabilities in VAEs pose a fundamental threat to watermark credibility.

\section{Threat Model}
\label{section:threatmodel}
% ----------------------------------------------------

\subsection{Attacker}

We consider the attackers composed of two collaborating roles within the image generation ecosystem: a malicious model supplier who deploys a closed-source LDM and adversarial users who access this model via a public service API.

\mypara{Attacker's Goal}
The malicious model supplier's objective is to outwardly satisfy regulatory requirements for watermark deployment while covertly weakening watermark enforcement through planting a backdoor into the encoder of VAE.
As for the adversarial users, they aim to remotely access powerful LDMs to generate arbitrary images and disseminate them in an untraceable manner by incorporating backdoor triggers distributed by the malicious supplier.
A realistic scenario is that a malicious AI company seeking to destabilize a region coordinates distributed users to generate and disseminate harmful content while evading watermark detection. 
By appearing compliant during routine watermark detection, such a service may avoid being blocked or sanctioned.

\mypara{Attacker's Capability}
The malicious supplier possesses the technical expertise and infrastructure required to develop powerful LDMs. 
Owing to their control over the model pipeline, the supplier can flexibly deploy any state-of-the-art semantic watermarking schemes. 
In contrast, adversarial users lack the resources and expertise to train or deploy LDMs, but can locally apply a provider-distributed trigger without specialized hardware or knowledge.

\mypara{Attacker's Knowledge}
The malicious supplier has full knowledge of the deployed LDM, including its architecture and parameters. 
However, the supplier does not know in advance which watermarking scheme will be required by regulators, the specific backdoor auditing mechanisms that may be applied prior to model release, or the concrete image content that adversarial users intend to generate.
Adversarial users, on the other hand, are assumed to have no knowledge of the model internals, watermarking schemes, or detection procedures. 
They interact with the system purely through the exposed API and rely on externally distributed triggers without understanding the underlying model behavior.

\subsection{Defender}

The defender is a trusted third-party authority, such as a model-hosting platform (e.g., HuggingFace~\cite{huggingface_hub}) or a governmental regulator, responsible for auditing and deploying LDMs.

\mypara{Defender's Goal}
Their primary goal is to ensure that each deployed LDM correctly embeds designated watermarks into generated images and supports reliable watermark detection.

\mypara{Defender's Capability \& Knowledge}
The defender possesses complete white-box access to the LDMs (including the U-Net, VAE, and text encoder), as well as full knowledge of the deployed watermarking scheme. 
While they are able to execute backdoor detection algorithms on the LDMs, they lack the knowledge of the specific trigger pattern.

% ----------------------------------------------------
\section{Methodology}
\label{section:method}
% ----------------------------------------------------

\begin{figure*}[t]
\centering
\includegraphics[width=0.99\linewidth]{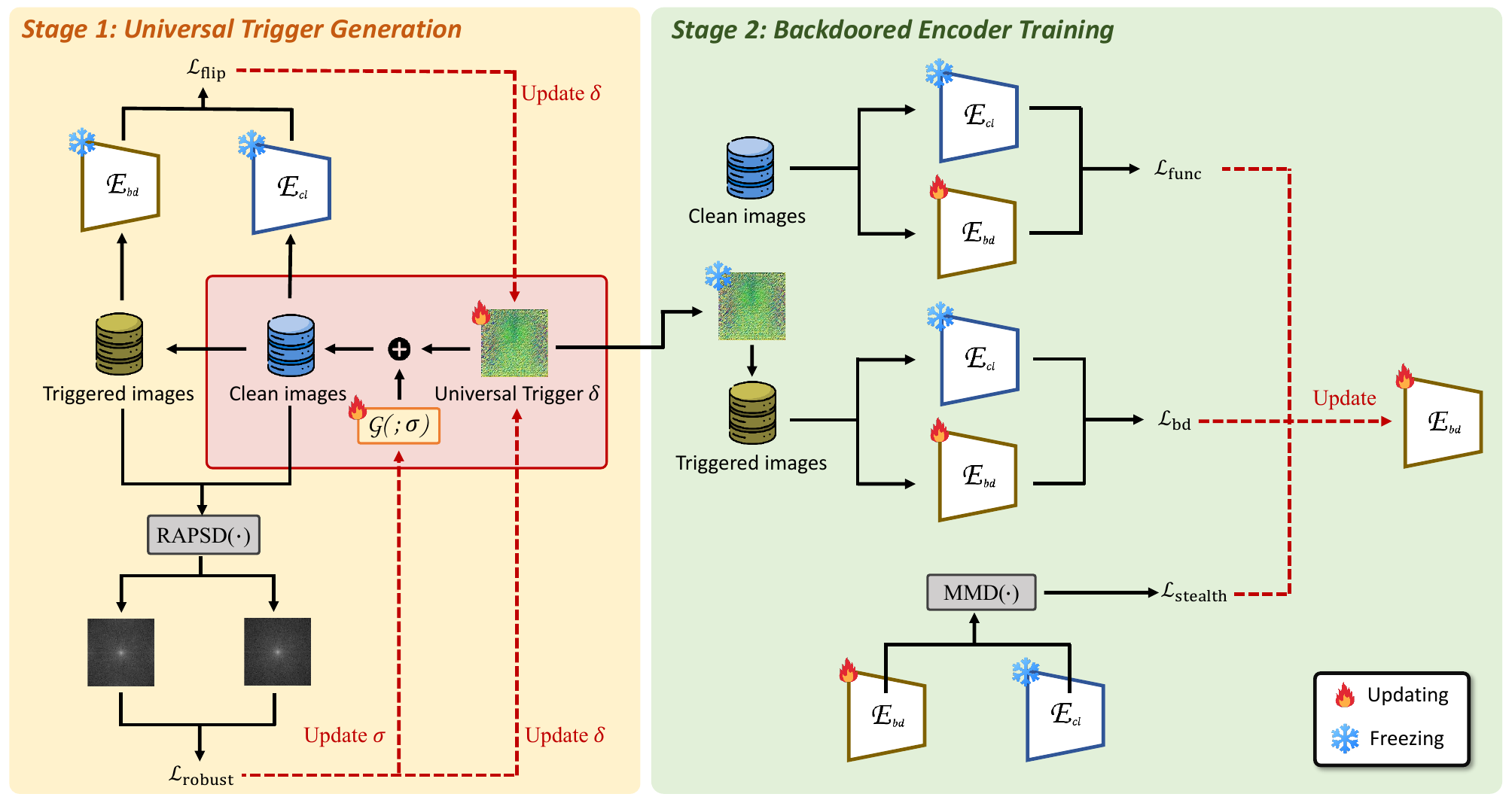}
\caption{\textbf{Overview of \mg.}
It proceeds in two stages.
\textbf{Stage 1} learns a universal, imperceptible, and robust trigger $\delta$ by optimizing a latent sign-flip objective $L_{\mathrm{flip}}$ while enforcing spectral concordance via the power spectrum regularization $L_{\mathrm{robust}}$.
\textbf{Stage 2} trains a backdoored VAE encoder $E_{{bd}}$ using a backdoor loss $L_{{bd}}$ under triggered inputs, a benign loss $L_{{func}}$ under clean images, while a parameter-alignment term $L_{\mathrm{stealth}}$ anchors $E_{{bd}}$ to the benign reference $E_{{cl}}$ for stealthiness.}
\label{fig:framework_fig}
\end{figure*}

\subsection{Overview}
\label{section:overview}

\mypara{Design Goal}
Our attack, \mg, is designed with explicit consideration of the practical requirements for a realistic and stealthy watermark evasion attack, and aims to achieve the following goals:
\begin{itemize}

\item \textbf{Selective Evasion:}
\mg aims to selectively evade semantic watermarking schemes by ensuring that only triggered watermarked images bypass detection, while benign watermarked images remain detectable.

\item \textbf{Backdoor Stealthiness:}
\mg seeks to optimize a stealth backdoored VAE encoder that evades detection by white-box auditors.
Stealthiness should be considered from three dimensions: input-space, parameter-space, and latent-space.

\item \textbf{Generation Utility Goal:}
\mg should maintain the core generative fidelity of LDMs, ensuring the standard text-to-image capability suffers no degradation in output quality.

\item \textbf{High Generalizability:}
The efficacy of \mg should be independent of both the LDM's architecture and the specific design of the semantic watermarking schemes.

\end{itemize}

\paragraph{Why Attack the VAE Encoder?}
The specific attacking goal of \mg is the VAE encoder, which is motivated by the structural asymmetry between the noise-to-image and image-to-noise pipelines.
As introduced in~\Cref{sec:ldm}, although both pipelines rely on the U-Net $\mathcal{U}(\cdot)$, they differ critically in how they interface with pixel space.
The noise-to-image generation process relies on the VAE decoder $\mathcal{D}(\cdot)$ to map the resulting latent into pixel space.
Therefore, by restricting the attack to the VAE encoder alone, the generation utility goal can be satisfied.

\subsection{\mg Framework}
\label{section:pipeline}
\mypara{High-level Design}
The \mg framework plants a backdoor into the VAE encoder using a sequential two-stage process: 
(1) \textit{Universal Trigger Generation} aims to optimize a robust and imperceptible backdoor trigger, and
(2) \textit{Backdoored Encoder Training} adapts the VAE encoder to enable selective evasion.
\Cref{fig:framework_fig} summarizes the workflow and the optimization targets in each stage.
Importantly, we exclusively use a dataset $\mathcal{D}_{cl}$ consisting of non-watermarked LDM-generated images during training, enabling the attack to generalize across different semantic watermarking schemes.
Formally, the goal of \mg is to generate a backdoored encoder $\mathcal{E}_{bd}$ from a benign encoder $\mathcal{E}_{cl}$ and a universal trigger $\delta$ as
$$\mathcal{E}_{bd},\delta \leftarrow \mathsf{GhostVAE}(\mathcal{E}_{cl},\mathcal{D}_{cl}),$$
where the malicious supplier retains $\mathcal{E}_{bd}$ for watermark detection while distributing the trigger $\delta$ to the colluded users. 

\begin{algorithm}[t]
\caption{Universal Trigger Generation}
\label{alg:mg_stage1_impl}
\DontPrintSemicolon
\LinesNumbered
\KwIn{
Clean dataset $\mathcal{D}_{cl}$; Backdoor encoder $\mathcal{E}_{bd}$; Perturbation budget $\varepsilon$;
Number of iteration steps $T_1$ and $T_2$; 
Learning rate $lr_1$ and $lr_2$
}
\KwOut{Universal trigger $\delta$}

Initialize trigger $\delta^{(1)}\sim\mathcal{N}(0,\mathbf{I}) $ and  blur intensity $\sigma^{(1)} \leftarrow 1$\;

\For{$t=1,\ldots,T_1$}{
$L_{\sigma}\leftarrow \mathbb{E}_{x \sim \mathcal{D}_{cl}}[\mathcal{L}_{\mathrm{robust}}(x,\delta^{(1)};\sigma^{(t)})]$\;
$\sigma^{(t+1)} \leftarrow \mathrm{Optimizer}(\sigma^{(t)};\ L_{\sigma},\ lr_1 )$\;
}
$\sigma \leftarrow \sigma^{(T_1)}$\;
\For{$t=1,\ldots,T_2$}{
\mbox{$L_{\delta}\leftarrow \mathbb{E}_{x \sim \mathcal{D}_{cl}}[
    \mathcal{L}_{\mathrm{flip}}(x,\delta^{(t)};\sigma)+\lambda \mathcal{L}_{\mathrm{robust}}(x,\delta^{(t)};\sigma)])$}\;
$\delta^{(t+1)} \leftarrow {\rm PGD}(\delta^{(t)}; L_{\delta},\varepsilon,lr_2)$\;
}
\Return Universal trigger $\delta\leftarrow\delta^{(T_2)}$\;
\end{algorithm}

\mypara{Stage 1: Universal Trigger Generation}
The pipeline of Stage 1 is shown in \Cref{alg:mg_stage1_impl}.
In this stage, we initialize the model as $\mathcal{E}_{bd}=\mathcal{E}_{cl}$ and freeze $\mathcal{E}_{bd}$ throughout this stage.
In brief, Stage 1 aims to optimize a universal trigger $\delta$ to possess two key characteristics: (i) the ability to flip the latent of the encoded image and (ii) robustness against common image corruptions, ensuring that the triggered image does not easily restore its watermark properties.
To achieve this objective, Stage 1 formulates the trigger generation process as the following optimization problem.
\begin{equation*}
\delta^*=\underset{\|\delta\|_{\infty}\le \varepsilon}{\arg\,\min}\;
\mathbb{E}_{x\sim\mathcal{D}_{cl}}
\Big[
\mathcal{L}_{\mathrm{flip}}(x,\delta)
\;+\;
\lambda\mathcal{L}_{\mathrm{robust}}(x,\delta)
\Big],
\label{eq:stage1_overview}
\end{equation*}
where $\lambda$ trades off attack efficacy against trigger robustness, and $\varepsilon$ specifies the $\ell_\infty$ perturbation budget.
We optimize the universal trigger under this constraint via Projected Gradient Descent (PGD)~\cite{DBLP:conf/iclr/MadryMSTV18}.
Next, we elaborate on the computational process of $\mathcal{L}_{\mathrm{robust}}$ and $\mathcal{L}_{\mathrm{flip}}$.

\begin{itemize}

\item \textbf{Robustness Goal:}
We use the power spectrum regularization technique from BlurGuard~\cite{kim2026blurguard} to enhance the robustness of our trigger $\delta$.
Specifically, BlurGuard is initially designed to improve the robustness of protective adversarial noise.
By adjusting the trigger's frequency spectrum, it avoids conspicuous high-frequency noise signatures.
Consequently, purification pipelines that rely on suppressing anomalous spectral components become less effective.
First of all, BlurGuard applies a Gaussian blur operator to $\delta$ via $\mathcal{G}(\delta;\sigma)$, aiming to attenuate high-frequency bands and effectively control the frequency spectrum of triggered images.
Here $\sigma$ is a learnable parameter that controls blur intensity.
Following~\cite{kim2026blurguard}, we decouple the optimization by first optimizing $\sigma$ (Line 2-4 in~\Cref{alg:mg_stage1_impl}).
Upon obtaining $\sigma$, BlurGuard further optimizes the trigger to pull the power spectrum of the triggered images back toward that of the original image through $\mathcal{L}_{\mathrm{robust}}$.
Therefore, $\mathcal{L}_{\mathrm{robust}}$ can be defined as
\begin{equation*}
\mathcal{L}_{\mathrm{robust}}(x,\delta;\sigma)
\;=\;
\Big\|
\log \frac{\mathrm{RAPSD}\!\big(F(x+\mathcal{G}(\delta;\sigma))\big)}{\mathrm{RAPSD}\!\big(F(x)\big)}
\Big\|_{\infty},
\label{eq:stage1_lrobust}
\end{equation*}
where $F(\cdot)$ computes the Fast Fourier Transform (FFT) and $\mathrm{RAPSD}(\cdot)$ calculates the radial average power spectral density.
The detailed calculation process of $\mathrm{RAPSD}(\cdot)$ is described in Appendix~\ref{appendix: rapsd}.

\item \textbf{Flipping Goal:}
We aim to introduce $\mathcal{L}_{\mathrm{flip}}$ to induce interference in the latent space of $\mathcal{E}_{bd}$.
Specifically, we encourage a systematic sign flip between the latent of a clean image $z=\mathcal{E}_{cl}(x)$ and the latent of its triggered counterpart $z'=\mathcal{E}_{bd}(x+\delta)$.
Note that in this stage, we still use the processed trigger $\mathcal{G}(\delta;\sigma)$.
Therefore, we define $\mathcal{L}_{\mathrm{flip}}$ as
\begin{equation*}
\label{eq:stage1_lflip}
\mathcal{L}_{\mathrm{flip}}(x,\delta;\sigma)
\;=\;
\big\|
\mathcal{E}_{bd}(x+\mathcal{G}(\delta;\sigma)) + \mathcal{E}_{cl}(x)
\big\|_2^2.
\end{equation*}

This design serves two purposes: 
(1) Sign-flipping introduces a stable interference in the latent inversion process that can evade the watermark detection process; 
(2) Sign-flipping preserves the latent distribution, improving the latent-space stealthiness.
Benefiting from this design, \mg can disrupt the detection results of watermarking schemes. 
Specifically, for Tree-Ring, because the Fourier transform is linear, the sign-flipped latent leads to a mismatched Fourier representation, which increases the distance from the target pattern $k$ and weakens detection.  
For Gaussian Shading and PRCMark, watermark detection depends on the sign of the recovered initial latent, which encodes the watermark codeword; sign flipping disrupts this codeword structure and causes the failure of watermark detection.
\end{itemize}

\mypara{Stage 2: Backdoored Encoder Training}
\mg further finetunes the parameters $\theta_{bd}$ of the model $\mathcal{E}_{bd}(\cdot;\theta_{bd})$ to adapt the trigger and achieve the parameter-space stealthiness.
The objectives of Stage 2 are threefold: preserving benign functionality, enforcing backdoor behavior, and enhancing backdoor stealthiness within the parameter space.
To this end, Stage 2 employs a multi-task learning approach to fine-tune the model parameters through the following optimization problem.
\begin{equation*}
\theta_{bd}^*=\underset{\theta}{\arg\min}\; [ \lambda_1 \mathcal{L}_{\rm func}(\cdot;\theta)
+ \lambda_2 \mathcal{L}_{\rm bd}(\cdot;\theta)
+ \lambda_3 \mathcal{L}_{\rm stealth}(\cdot;\theta)],
\end{equation*}
where hyperparameters $\lambda_1$, $\lambda_2$, and $\lambda_3$ control the relative importance of each optimization goal.
We now proceed to introduce each objective separately.

\begin{itemize}

\item \textbf{Functionality Goal: }
Given a clean image $x$, \mg expects its output latent to be consistent with that of the benign encoder. 
Therefore, \mg uses the latent generated by the original frozen $\mathcal{E}_{cl}(\cdot)$ as the supervisory signal to optimize the following loss function:
\begin{equation*}
\mathcal{L}_{\rm func}(x;\theta_{bd})=
\mathbb{E}_{x\sim\mathcal{D}_{cl}}
[||\mathcal{E}_{bd}(x;\theta_{bd}) - \mathcal{E}_{cl}(x)||_2^2].
\end{equation*}

\item \textbf{Evading Goal: }For the triggered inputs, the encoder needs to continue updating to enhance the attack capability of the trigger, ensuring that the trigger keeps satisfying the sign-flip constraint for the latents.
\begin{equation*}
\mathcal{L}_{\rm bd}(x;\theta_{bd})=
\mathbb{E}_{x\sim\mathcal{D}_{cl}}
[||\mathcal{E}_{bd}(x+\delta;\theta_{bd}) + \mathcal{E}_{cl}(x)||_2^2].
\end{equation*}

\item \textbf{Stealthiness Goal: }
\mg enforces the stealthiness in the parameter space through a channel-level distribution alignment regularizer.
This design is motivated by recent findings showing that backdoor behavior often concentrates in a small subset of neurons, yielding detectable parameter or activation anomalies~\cite{DBLP:conf/ccs/Xu0KP25,Zhai_2025_ICCV}.
To prevent this, \mg enforces parameter-space similarity by minimizing the distributional discrepancy between the backdoored and benign weights using Maximum Mean Discrepancy (MMD)~\cite{10.5555/2188385.2188410}.
By anchoring each layer of the backdoored encoder to its benign counterpart, this regularizer suppresses abnormal channel-wise drift.
Specifically, for each convolution layer $k$ of $\theta_{bd}$, let the parameter weights be $W^{(k)}\in\mathbb{R}^{C_{\text{out}}\times C_{\text{in}}\times h\times w}$, where $C_{\text{out}}$ and $C_{\text{in}}$ are the numbers of output and input channels, and
$h\times w$ is the spatial size of the convolution kernel.
We flatten the last three dimensions, obtaining two parameter sets
$W_{bd}^{(k)}, W_{{cl}}^{(k)}\in\mathbb{R}^{C_{\mathrm{out}}\times C}$,
where $C=C_{\mathrm{in}}hw$.
Equivalently, each set contains $C_{\mathrm{out}}$ channel-wise vectors in $\mathbb{R}^{C}$:
$W_{bd}^{(k)}=\{x^{(k)}_{{bd},i}\}_{i=1}^{C_{\mathrm{out}}}$ and
$W_{{cl}}^{(k)}=\{x^{(k)}_{{cl},j}\}_{j=1}^{C_{\mathrm{out}}}$.
We can then compute the MMD value at layer $k$ as
\[
\begin{aligned}
MMD(W^{(k)}_{bd},W^{(k)}_{cl})=
&\frac{1}{n_k^2}\sum_{i=1}^{n_k}\sum_{i'=1}^{n_k}
\kappa\!\left(x^{(k)}_{bd,i},x^{(k)}_{bd,i'}\right) \\
&+
\frac{1}{m_k^2}\sum_{j=1}^{m_k}\sum_{j'=1}^{m_k}
\kappa\!\left(x^{(k)}_{cl,j},x^{(k)}_{cl,j'}\right) \\
&-\frac{2}{n_k m_k}\sum_{i=1}^{n_k}\sum_{j=1}^{m_k}
\kappa\!\left(x^{(k)}_{bd,i},x^{(k)}_{cl,j}\right),
\end{aligned}
\]
with $n_k=m_k=C_{\text{out}}$ and the kernel is
\[
\kappa(x,y)=\sum_{s=1}^{S}\exp\!\left(-\frac{\lVert x-y\rVert_2^2}{2\sigma_{k,s}^2}\right),
\]

where S is the number of kernel scales and $\sigma_{k,s}$ are fixed multipliers. 
The final loss function is obtained by averaging the values from all matched layers.
\begin{equation*}
\mathcal{L}_{\rm stealth}=\mathbb{E}_{k\in\mathcal{K}}[MMD(W_{bd}^{(k)},W_{cl}^{(k)})],
\end{equation*}
where $\mathcal{K}$ denotes the set of matched convolution layers.

\end{itemize}

% ----------------------------------------------------
\section{Experiments}
\label{section:exp}
% ----------------------------------------------------

\subsection{Experimental Setup}

\begin{table*}[t]
\centering
\caption{Details of diffusion models and their used VAEs.}
\label{tab:diffusion_models}
\setlength{\tabcolsep}{10pt}
\begin{tabular}{lccc}
\toprule
Model & Company & VAE   & VAE's description \\
\midrule
SD-2.1~\cite{Rombach2021HighResolutionIS}
& Stability AI
& \texttt{stabilityai/sd-vae-ft-mse}~\cite{sdvaemseft}
& Fine-tune of the KL-f8~\cite{Rombach2021HighResolutionIS} \\
SD-XL~\cite{podell2024sdxl} 
& Stability AI
& \texttt{stabilityai/sdxl-vae}~\cite{sdxlvae}
& Retrained from scratch~\cite{podell2024sdxl} \\
Flux-1.0~\cite{labs2025flux1kontextflowmatching}
& Black Forest Labs
& \texttt{black-forest-labs/FLUX-VAE}~\cite{flux2024}
& Retrained from scratch~\cite{labs2025flux1kontextflowmatching} \\
\bottomrule
\end{tabular}
\end{table*}

\mypara{Diffusion Models}
As summarized in~\Cref{tab:diffusion_models}, we evaluate \mg on three representative open-source LDMs.
These LDMs rely on different VAE designs, which allows us to test whether \mg generalizes across models.
For all LDMs, we generate images using their default inference configurations.

\begin{itemize}

\item \textbf{SD-2.1~\cite{Rombach2021HighResolutionIS}:} 
stable-diffusion-2-1-base (SD-2.1) is a widely used LDM for $512\times512$ text-to-image generation.
It adopts the standard LDM pipeline with a U-Net denoiser and a KL-regularized VAE.
We use its default VAE \texttt{stabilityai/sd-vae-ft-mse} for encoding and decoding.

\item \textbf{SD-XL~\cite{podell2024sdxl}:} 
stable-diffusion-xl-base-1.0 (SD-XL) is a higher-capacity LDM that substantially improves visual fidelity and prompt adherence over SD-2.1.
It uses a different VAE (\texttt{stabilityai/sdxl-vae}) and a larger denoising backbone, making it a strong testbed for assessing whether the proposed encoder backdoor remains effective under a more powerful generation stack.

\item \textbf{Flux-1.0~\cite{labs2025flux1kontextflowmatching}:} 
Flux.1-dev (Flux-1.0) represents a newer generation paradigm based on transformer-style backbones and flow-matching objectives.
It operates in the latent space of a high-capacity convolutional autoencoder and uses a 16-channel latent representation produced by the \texttt{black-forest-labs/FLUX-VAE}, which differs markedly from the 4-channel VAEs used in SD-2.1 and SD-XL.
\end{itemize}

\mypara{Datasets}
To construct $\mathcal{D}_{cl}$ of \mg, we sample text prompts from the SDP dataset~\cite{gustavosta_sd_prompts} and generate 8,000 unwatermarked images.
For evaluation, we sample a disjoint set of 1,000 prompts to generate watermarked images with different LDMs and different watermarking schemes.

\mypara{Metrics}
We evaluate the effectiveness of \mg from two aspects: watermark detection and visual fidelity.
\begin{itemize}
\item \textbf{Watermark Detection.}
We report the true positive rate (TPR) and the false positive rate (FPR) of the watermarking schemes.
TPR is the fraction of watermarked images that are correctly verified as watermarked,
whereas FPR is the fraction of unwatermarked images that are incorrectly flagged as watermarked.
\item \textbf{Attack Success Rate (ASR).}
ASR measures the fraction of watermarked images that successfully evade detection after adding the trigger.
\item \textbf{Image Quality.}
To quantify the perceptual impact of $\delta_{trigger}$, we report Peak Signal-to-Noise Ratio (PSNR, $\uparrow$), Structural Similarity Index Measure (SSIM, $\uparrow$), and Learned Perceptual Image Patch Similarity (LPIPS, $\downarrow$) between each image and its triggered counterpart.
\end{itemize}

\mypara{Hyperparameters of \textbf{\mg}}
In Stage~1 of \mg, we set $\lambda=50.0$, $T_1=T_2=10$, $lr_1=lr_2=10^{-4}$, and a default perturbation budget of $\varepsilon=8/255$. 
In Stage 2, we fine-tune the VAE encoder for $T=4$ epochs using AdamW~\cite{Loshchilov2017DecoupledWD} with learning rate $\eta=10^{-4}$.
Moreover, we set $\lambda_{1}=\lambda_{2}=1.0$ and $\lambda_{3}=50.0$.
All experiments are conducted on a server with 8 NVIDIA A800 GPUs.

\mypara{Semantic Watermarking Schemes}
We implant watermarks on images with the size of $512\times512$ for all schemes.
For Tree-Ring, we set the radius of pattern $k$ to $4$ for SD-2.1 and SD-XL, and to $6$ for FLUX-1.0; the masked region is placed in the last latent channel.
For Gaussian Shading, we set the message length to $k=256$ and use repetition factor $s=64$ for SD-2.1 and SD-XL, increasing to $s=256$ for FLUX-1.0.
For PRCMark, we set $t=3$.
The thresholds setting for these watermarking schemes are shown in Appendix~\ref{appendix: threshold-setting}.

\subsection{Performance of \mg}
We evaluate the effectiveness of \mg along two complementary dimensions: (i) its ability to invalidate watermark detection in the presence of a trigger and (ii) its ability to preserve the benign functionality of the VAE encoder on non-triggered inputs.

\mypara{Attack Performance}
\Cref{tab:attack_results_asr_main} summarizes the attack success rate (ASR) across three different LDMs and three representative watermarking schemes.
Overall, \mg achieves consistently high ASR with an imperceptible trigger.
In particular, PRCMark detection is almost completely neutralized across models, reaching a $100\%$ ASR on SD-2.1 and exceeding $95\%$ on both SD-XL and FLUX-1.0.
Gaussian Shading and Tree-Ring are also strongly affected, with ASR remaining high across different LDMs.
These results demonstrate that \mg reliably disrupts watermark verifiability while generalizing across different models and watermarking schemes.

\begin{table}[t]
\centering
\caption{Attack performance of \mg.}
\label{tab:attack_results_asr_main}
\setlength{\tabcolsep}{14pt}
\begin{tabular}{llc}
\toprule
Model & Watermark & ASR \\
\midrule
\multirow{3}{*}{SD-2.1}
& Tree-Ring & 88.2\% \\
& Gaussian Shading      & 86.0\% \\
& PRCMark     & 100.0\%  \\
\midrule
\multirow{3}{*}{SD-XL}
& Tree-Ring & 97.1\% \\
& Gaussian Shading      & 93.8\% \\
& PRCMark    & 99.0\% \\

\midrule
\multirow{3}{*}{Flux-1.0}
& Tree-Ring & 99.1\% \\
& Gaussian Shading      & 92.5\% \\
& PRCMark   & 96.0\% \\

\bottomrule
\end{tabular}
\end{table}

\begin{table}[t]
\centering
\caption{Function maintenance comparison between the base VAE and \mg (\%).}
\label{tab:attack_results_tpr_fpr_main}
\setlength{\tabcolsep}{4pt}
\begin{tabular}{ll rr rr}
\toprule
\multirow{2}{*}{Model} & \multirow{2}{*}{Watermark} &
\multicolumn{2}{c}{Base VAE} &
\multicolumn{2}{c}{\mg} \\
\cmidrule(lr){3-4}\cmidrule(lr){5-6}
& & TPR & FPR & TPR & FPR \\
\midrule
\multirow{3}{*}{SD-2.1}
& Tree-Ring       & 96.8 & 0.0 & 93.5  & 0.0 \\
& Gaussian Shading & 95.9 & 0.0 & 98.7  & 0.0 \\
& PRCMark         & 100.0 & 0.0 & 100.0 & 0.0 \\
\midrule
\multirow{3}{*}{SD-XL}
& Tree-Ring       & 98.1 & 0.3 & 90.1  & 0.1 \\
& Gaussian Shading & 98.6 & 0.0 & 87.0  & 0.0 \\
& PRCMark         & 94.0 & 0.0 & 86.4  & 0.0 \\
\midrule
\multirow{3}{*}{Flux-1.0}
& Tree-Ring       & 94.5 & 0.0 & 94.0  & 0.0 \\
& Gaussian Shading & 99.8 & 0.0 & 100.0 & 0.0 \\
& PRCMark         & 99.7 & 0.0 & 99.6  & 0.0 \\
\bottomrule
\end{tabular}
\end{table}

\begin{figure}[t]
\centering
\begin{minipage}{0.32\linewidth}
    \centering
    \includegraphics[width=\linewidth]{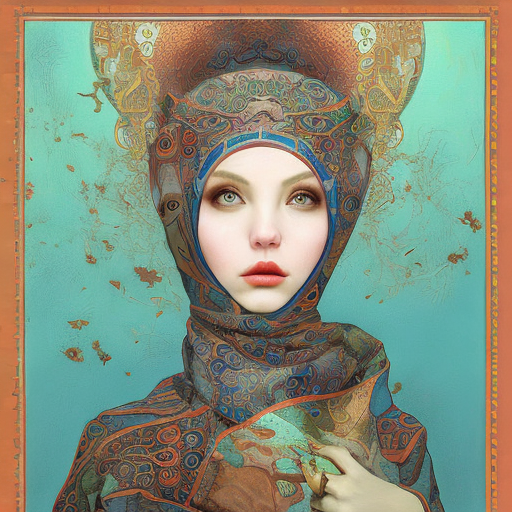}
    \small (a) Input
\end{minipage}
\hfill
\begin{minipage}{0.32\linewidth}
    \centering
    \includegraphics[width=\linewidth]{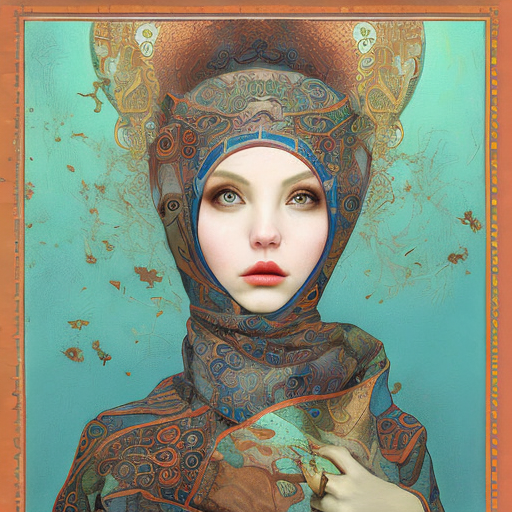}
    \small (b) Base VAE
\end{minipage}
\hfill
\begin{minipage}{0.32\linewidth}
    \centering
    \includegraphics[width=\linewidth]{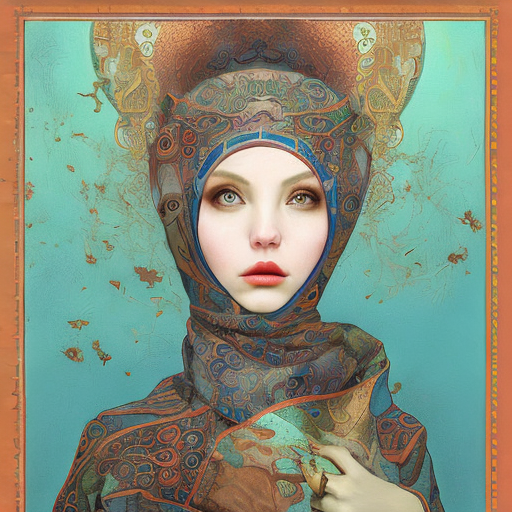}
    \small (c) \mg
\end{minipage}
\caption{Qualitative comparison of reconstruction quality. The base VAE and \mg produce visually similar reconstructions on non-triggered inputs. The LDM is SD-2.1.}
\label{fig:benign_reconstruction}
\end{figure}

\mypara{Function Maintenance}Beyond the attack effectiveness, \mg must preserve benign behavior on non-triggered inputs.
\Cref{tab:attack_results_tpr_fpr_main} reports the TPR on clean watermarked images and the FPR on unwatermarked images between the base VAE and \mg.
Overall, \mg largely maintains the benign detection capability.
The only notable deviation appears on SD-XL with Gaussian Shading, where TPR decreases by above 10\%.
However, the drop is acceptable because the resulting TPR is still high, and this reduction does not affect \mg's FPR.
This confirms that \mg selectively disrupts detection only when the trigger is present, without inducing spurious detections on unwatermarked content.
We further assess benign reconstruction quality on 200 non-triggered images. 
The mean PSNR decreases only marginally, from 35.03dB for the base VAE to 34.98dB for \mg, indicating negligible degradation in encoder-dependent reconstruction behavior.
A qualitative comparison is provided in \Cref{fig:benign_reconstruction}.

\subsection{Ablation Study}

\mypara{The Necessity of Stage 1}
Recall that the pipeline of \mg involves two stages: Stage 1 seeks to optimize a universal robust trigger and Stage 2 finetunes the VAE encoder to plant the backdoor.
We first assess the necessity of the trigger robustness design in Stage 1. 
To this end, we test the trigger under several input-space image perturbations, which are also representative input-space defenses and will be analyzed in detail in \Cref{sec:inputdefense}.
As shown in \Cref{tab:input_mitigation_robustness}, the trigger optimized without the Stage 1 robustness design becomes substantially less reliable under several input-space perturbations, whereas the full \mg consistently preserves a high ASR.
These results demonstrate that the trigger robustness design is essential for preserving stable watermark evasion under image corruption and purification.

\mypara{The Necessity of Stage 2}
To prove that merely optimizing a universal trigger is challenging, we initiate the attack by removing Stage 2 in this part.
This setting isolates whether a purely input-space perturbation can reliably evade watermark detection.
The results in \Cref{tab:noise_only_ablation} demonstrate that the exclusion of Stage 2 leads to a sharp decline in attack performance.
In contrast, the full \mg attains $100\%$ ASR under all three budgets. 
These results confirm that the effectiveness of \mg relies on jointly shaping the encoder’s response to the trigger through backdoor training.
\begin{table}[t]
\centering
\caption{Discussion on the necessity of Stage 2.
The LDM is SD-2.1 and the watermarking scheme is PRCMark.}
\label{tab:noise_only_ablation}
\setlength{\tabcolsep}{3.5pt}
\begin{tabular}{c cc}
\toprule
$\varepsilon$ & \mg (w/o Stage 2) & \mg (w/ Stage 2) \\
\midrule
$2/255$ & 3.6\% & 100.0\% \\
$4/255$ & 14.8\% & 100.0\% \\
$8/255$ & 31.6\% & 100.0\% \\
\bottomrule
\end{tabular}
\end{table}

\begin{table}[t]
\centering
\caption{Attack performance of \mg under tighter perturbation budgets.}
\label{tab:attack_results_asr_ablation_eps}
\setlength{\tabcolsep}{5pt}
\begin{tabular}{ll cc}
\toprule
Model & Watermark & $\varepsilon=2/255$ & $\varepsilon=4/255$ \\
\midrule
\multirow{3}{*}{SD-2.1}
& Tree-Ring & 88.5\% & 88.9\% \\
& Gaussian Shading      & 80.9\% & 83.2\% \\
& PRCMark     & 100.0\%  & 100.0\%  \\
\midrule
\multirow{3}{*}{SD-XL}
& Tree-Ring & 97.5\% & 96.5\% \\
& Gaussian Shading      & 91.3\% & 91.8\% \\
& PRCMark     & 97.5\% & 98.3\% \\
\midrule
\multirow{3}{*}{Flux-1.0}
& Tree-Ring & 86.9\% & 95.8\% \\
& Gaussian Shading      & 84.8\% & 90.6\% \\
& PRCMark     & 85.3\% & 92.8\% \\
\bottomrule
\end{tabular}
\end{table}

\begin{table}[t]
\centering
\caption{Function maintenance of \mg under tighter perturbation budgets.}
\label{tab:attack_results_tpr_fpr_ablation_eps}
\setlength{\tabcolsep}{3pt}
\begin{tabular}{ll cc cc}
\toprule
\multirow{2}{*}{Model} & \multirow{2}{*}{Watermark} &
\multicolumn{2}{c}{$\varepsilon=2/255$} &
\multicolumn{2}{c}{$\varepsilon=4/255$} \\
\cmidrule(lr){3-4}\cmidrule(lr){5-6}
& & TPR & FPR & TPR & FPR \\
\midrule
\multirow{3}{*}{SD-2.1}
& Tree-Ring & 93.1\% & 0.0\% & 93.1\% & 0.0\% \\
& Gaussian Shading      & 98.8\% & 0.0\% & 99.0\% & 0.0\% \\
& PRCMark     & 100.0\%  & 0.0\% & 100.0\%  & 0.0\% \\
\midrule
\multirow{3}{*}{SD-XL}
& Tree-Ring & 89.9\% & 0.0\% & 89.1\% & 0.0\% \\
& Gaussian Shading      & 85.9\% & 0.0\% & 86.8\% & 0.0\% \\
& PRCMark     & 84.6\% & 0.0\% & 85.9\% & 0.0\% \\
\midrule
\multirow{3}{*}{Flux-1.0}
& Tree-Ring & 94.4\% & 0.0\% & 91.2\% & 0.0\% \\
& Gaussian Shading      & 100.0\%  & 0.0\% & 99.9\% & 0.0\% \\
& PRCMark     & 99.7\% & 0.0\% & 99.6\% & 0.0\% \\
\bottomrule
\end{tabular}
\end{table}

\begin{table}[t]
\centering
\caption{Image quality after adding the optimized noise trigger under different perturbation budgets.
We summarize the distribution using the mean $\pm$ standard deviation over 1,000 images.}
\label{tab:trigger_quality_metrics}
\setlength{\tabcolsep}{3.2pt}
\renewcommand{\arraystretch}{1.05}
\begin{tabular}{rccc}
\toprule
\textbf{Metric} & $\varepsilon=2/255$ & $\varepsilon=4/255$ & $\varepsilon=8/255$ \\
\midrule
PSNR $\uparrow$   & $43.82 \pm 0.07$ & $38.47 \pm 0.09$ & $33.96 \pm 0.11$ \\
SSIM $\uparrow$   & $0.984 \pm 0.004$ & $0.955 \pm 0.012$ & $0.904 \pm 0.024$ \\
LPIPS $\downarrow$& $0.030 \pm 0.031$ & $0.083 \pm 0.059$ & $0.162 \pm 0.083$ \\
\bottomrule
\end{tabular}
\end{table}

\subsection{Parameter Sensitivity}
\mypara{Impact of Perturbation Budget $\varepsilon$}
\label{sec:ablation_eps}
We ablate the perturbation budget by tightening the perturbation constraint to $\varepsilon\in\{2/255,4/255\}$.
The results of ASR are shown in \Cref{tab:attack_results_asr_ablation_eps}.
We can observe that even at $\varepsilon=2/255$, \mg remains highly effective: \mg perfectly evades the PRCMark on SD-2.1 with a 100.0\% ASR and stays strong on SD-XL, achieving 97.5\% ASR. 
Also, ASRs on Gaussian Shading and Tree-Ring typically remain above 80\%, demonstrating strong attack performance.
Increasing the budget to $\varepsilon=4/255$ further improves ASR across nearly all settings.
Moreover, \Cref{tab:attack_results_tpr_fpr_ablation_eps} reports TPR and FPR at $\varepsilon=2/255$ and $4/255$.
Across these models and watermarking schemes, TPR remains high, and FPR stays at $0.0\%$.
Therefore, shrinking the trigger budget has almost no impact on the attacking performance of \mg and the encoder’s benign detection behavior.
We further evaluate the impact of the perturbation budget on the visual quality.
\Cref{tab:trigger_quality_metrics} reports PSNR/SSIM/LPIPS between each image and its triggered counterpart.
These results indicate that \mg achieves watermark evasion without introducing conspicuous visual artifacts, keeping the attack practical and stealthy.

\mypara{Impact of Loss Weights}
We evaluate the sensitivity of \mg to its loss weights on SD-2.1 with PRCMark by varying one coefficient at a time while keeping the others fixed. Specifically, we study the trigger robustness weight $\lambda$ (30, 70, 90, 110), the functionality preserving weight $\lambda_1$ (2, 3, 5, 8), the backdoor enforcing weight $\lambda_2$ (2, 3, 5, 8), and the parameter stealth weight $\lambda_3$ (30, 70, 90, 110), resulting in 16 configurations in total. Across all tested settings, \mg consistently achieves an ASR of 100\%, showing that the trigger reliably disables watermark detection once activated. Meanwhile, the TPR on clean watermarked images remains at 99\%--100\%, and the FPR stays at 0\%. These results indicate that, under the standard evaluation setting, \mg maintains stable attack effectiveness and benign detection behavior across a reasonably broad range of loss weight choices.

\subsection{Compare with Watermark Removal Attacks}
\label{sec:55dis_removal}
\mypara{Baseline Attacks}
While the objective of \mg is to evade detection by watermarking schemes, similarly, several advanced watermark removal attacks have recently been proposed. 
We further compare \mg with three representative watermark removal attacks: \textbf{Embedding attack}~\cite{DBLP:conf/icml/AnDRAXDZMWGH24}, \textbf{Imprint attack}~\cite{DBLP:conf/cvpr/0025LTFQ25}, and \textbf{Regeneration attack}~\cite{DBLP:conf/nips/ZhaoZSVGKVWL24}.
In our evaluation, we follow a realistic black-box setting by using SD-1.5 as the proxy model to attack SD-2.1 and evaluate all methods on 100 PRCMark-based watermarked images.
We set the number of optimization steps for the Imprint attack to $T_{\text{imprint}}=30$ and the number of denoising steps for the Regeneration attack to $T_{\text{reg}}=200$.

\mypara{Evaluation Results}
The corresponding results are shown in~\Cref{tab:removal_attack_result}.
Initially, we compare the attack performance. \mg achieves a 100\% ASR, while Imprint also attains a 97\% ASR. 
In contrast, the Embedding and Regeneration methods only achieve 80\% and 38\% ASR, respectively. 
This disparity indicates that their attack performance is unstable, even under conditions of high computational cost.
Meanwhile, we notice that these removal attacks significantly degrade the quality of the watermarked images. 
In contrast, under the evaluation of three different metrics, \mg consistently maintains the best image quality. 
For instance, the PSNR of \mg is 33.96dB, while those of Imprint and Regeneration are only 23.66dB and 21.55dB, respectively.
Finally, we assess the time cost associated with the attacks. 
For \mg, we consider only the time required by a malicious user to apply the trigger to a watermarked image. 
The results indicate that the computational overhead of \mg is nearly negligible, whereas Imprint requires 173.36s to generate a single successfully attacked image.
We further provide an ablation on the iteration budget $T_{imprint}
$ for Imprint attack and denoising steps $T_{reg}$ for Regeneration attack in \Cref{appendix:removal_steps}.
Collectively, these results demonstrate that evading watermark detection via a backdoor imposes lower capability requirements on malicious adversaries while representing a higher-level threat.

\begin{table}[t]
\centering
\caption{Comparison with Embedding, Imprint, and Regeneration attacks under a black-box setting (SD-1.5 $\rightarrow$ SD-2.1), with $\epsilon_{\text{embedding}} = 8/255$, $T_{\text{imprint}} = 30$, and $T_{\text{reg}} = 200$.}

\label{tab:removal_attack_result}
\setlength{\tabcolsep}{2.5pt}
\begin{tabular}{rcccc}
\toprule
Metrics & Embedding & Imprint & Regeneration & \mg \\
\midrule
ASR   $\uparrow$       & 80.0\%  & 97.0\% & 38.0\% & 100.0\% \\
PSNR $\uparrow$   & 31.46   & 23.66  & 21.55  &33.96 \\
SSIM $\uparrow$   & 0.82    & 0.64      & 0.58     & 0.90 \\
LPIPS $\downarrow$ & 0.24    & 0.16      & 0.24     & 0.16 \\
Time  $\downarrow$        & 5.83s    & 173.36s     & 0.38s   & \num{5e-5}s\\
\bottomrule
\end{tabular}
\end{table}

\section{Defenses}
\label{sec:defenses}
The defenses against backdoored neural networks have remained a prominent research focus~\cite{DBLP:journals/ftsec/MaGWWWSDXCZHLWZZBLWQZHL25, DBLP:journals/ijcv/WuCZZWYZWLS25,8835365,9338311}.
We demonstrate the stealthiness of \mg across three dimensions: input-space, parameter-space, and latent-space, spanning a total of 17 defenses. 
Note that for certain classical defense methods, such as ANP~\cite{DBLP:conf/iclr/ZengCPM0J22}, that were originally proposed on classifiers, we have adapted them to operate within the VAE scenario.

\subsection{Input-space Defenses}
\label{sec:inputdefense}
\mypara{Image Corruption}
Once the malicious supplier releases triggered images, their dissemination across networks inevitably subjects them to real-world distortions.
Here, we evaluate five common environmental corruptions: additive Gaussian noise, JPEG compression, Resize, Center crop, Rotation.
For Gaussian noise, $\sigma$ denotes the standard deviation of the added noise.
For JPEG, $Q$ denotes the compression quality factor.
For Resize, $S$ denotes the resizing factor.
For Center crop, $R$ denotes the retained crop ratio.
For Rotation, $A$ denotes the rotation angle.
The results are shown in \Cref{tab:input_mitigation_robustness}.
These perturbations affect watermark detection to different degrees. 
Some strong geometric transformations, such as Center crop and larger Rotations, substantially degrade TPR, resulting in their high $ASR_1$ and $ASR_2$. 
By contrast, perturbations such as JPEG compression and Resize largely preserve watermark verifiability in our settings; the consistently high $ASR_2$ under these conditions more directly demonstrates the robustness of our trigger design.
In summary, optimizing the trigger \emph{without} the robustness goal mentioned in \Cref{section:pipeline} is noticeably more fragile, whereas the full \mg consistently preserves a stable high attack success.

\mypara{Perturbation Purification}
Furthermore, we consider the scenario where external users implement adaptive adversarial purification defenses. 
An example is the application of techniques like DiffPure~\cite{DBLP:conf/icml/NieGHXVA22} to regenerate images to neutralize the adversarial perturbations.
$T$ is the number of denoising steps used for purification.
As reported in \Cref{tab:input_mitigation_robustness},
DiffPure increasingly weakens watermark verifiability as $T$ grows. 
Nevertheless, the trigger equipped with our robustness design remains highly effective even with purification at $T=5$ and $T=10$.

\mypara{Social Media Pipeline}
We also evaluate robustness under social media pipeline by posting/reposting triggered images on X (formerly Twitter)~\cite{X} and downloading them for detection. As shown in \Cref{tab:input_mitigation_robustness}, both settings preserve 100\% TPR and 100\% $ASR_2$, indicating that the tested X pipeline does not influence our attack.

\begin{table}[t]
\centering
\caption{Trigger robustness against input-space disturbances. 
$ASR_1$ is obtained with a trigger optimized without the robustness goal, while $ASR_2$ corresponds to the full \mg.
The results are evaluated on SD-2.1 for 100 images, and the watermarking scheme is PRCMark.
}
\label{tab:input_mitigation_robustness}
\setlength{\tabcolsep}{4pt}
\renewcommand{\arraystretch}{1.15}
\begin{tabular}{llcccc}
\toprule
Defense & Setting & TPR & $ASR_1$ & $ASR_2$ \\
\midrule
\multirow{3}{*}{Gaussian Noise} 
& $\sigma=0.01$ & 100\% & 100\% & 100\% \\
& $\sigma=0.02$ & 95\% & 100\% & 100\% \\
& $\sigma=0.03$ & 86\% & 100\% & 100\% \\
\midrule
\multirow{3}{*}{JPEG}
& $Q=80$ & 98\% & 99\% & 100\% \\
& $Q=60$ & 93\% & 41\% & 100\% \\
& $Q=40$ & 86\% & 32\% & 100\% \\
\midrule
\multirow{3}{*}{Resize}
& $S$ = 0.95 & 100\%& 100\% & 100\% \\
& $S$ = 0.90 & 100\%& 99\% & 100\% \\
& $S$ = 0.85 & 100\%& 97\% & 99\% \\
\midrule
\multirow{3}{*}{Center Crop}
& $R$ = 0.98 & 2\% & 100\%& 100\% \\
& $R$ = 0.95 & 0\% & 100\%& 100\% \\
& $R$ = 0.90 & 0\% & 100\%& 100\% \\
\midrule
\multirow{3}{*}{Rotation}
& $A$ = 0.5$^\circ$ & 99\%& 95\% & 100\% \\
& $A$ = 1.0$^\circ$ & 20\%& 100\% & 100\% \\
& $A$ = 1.5$^\circ$ & 0\% & 100\%& 100\% \\
\midrule
\multirow{3}{*}{DiffPure~\cite{DBLP:conf/icml/NieGHXVA22}}
& $T=5$  & 87\% & 29\% & 100\% \\
& $T=10$ & 72\% & 46\% & 100\% \\
& $T=15$ & 53\% & 54\% & 99\% \\
\midrule
\multirow{2}{*}{Social Media Pipeline}
& X post & 100\%& 100\% & 100\% \\
& X repost & 100\%& 100\% & 100\% \\
\bottomrule
\end{tabular}
\end{table}

\subsection{Parameter-space Defenses}
\label{sec: paradefense}

\mypara{Model Pruning}
\begin{figure*}[t]
\centering
\begin{subfigure}[t]{0.33\linewidth}
\centering
\includegraphics[width=\linewidth]{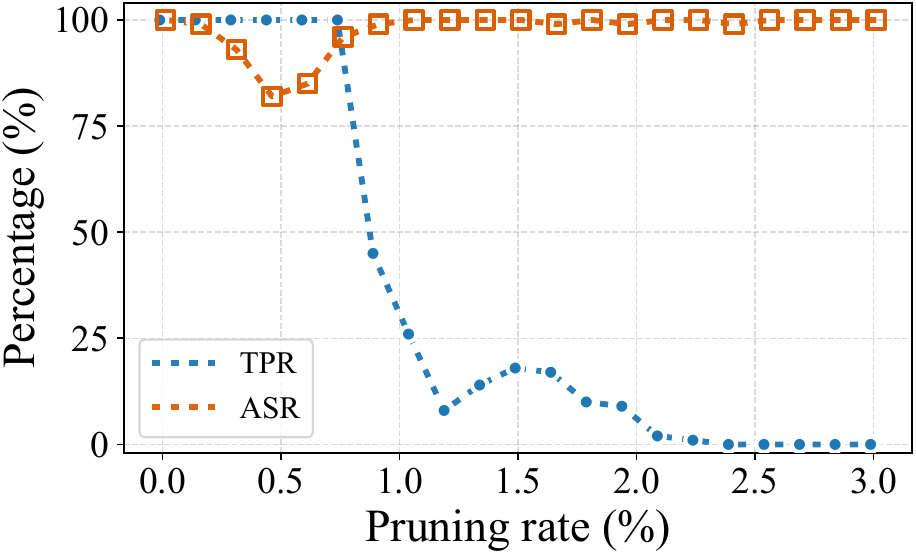}
\caption{Iterative channel-space structured pruning.}
\label{fig:prc_prune}
\end{subfigure}
\hfill
\begin{subfigure}[t]{0.33\linewidth}
\centering
\includegraphics[width=\linewidth]{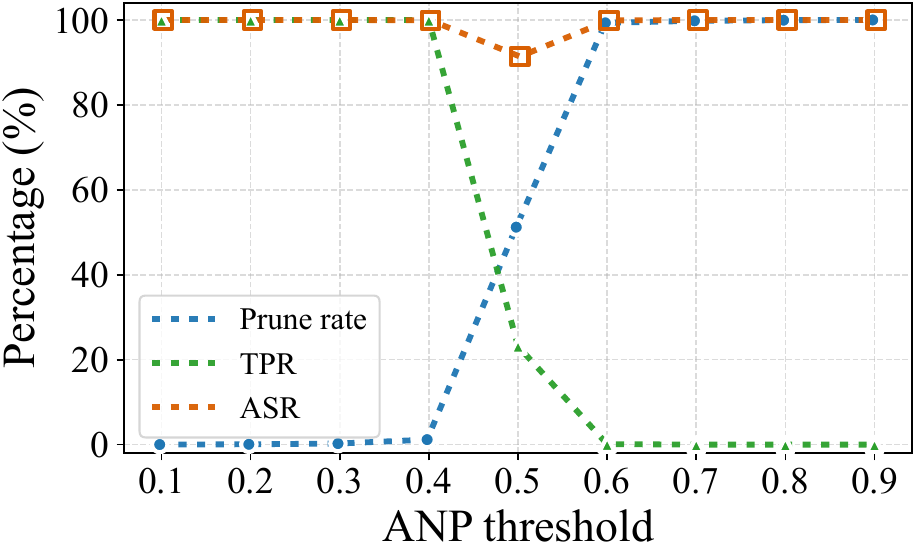}
\caption{ANP results with threshold sweeping.}
\label{fig:prc_anp}
\end{subfigure}
\hfill
\begin{subfigure}[t]{0.33\linewidth}
\centering
\includegraphics[width=\linewidth]{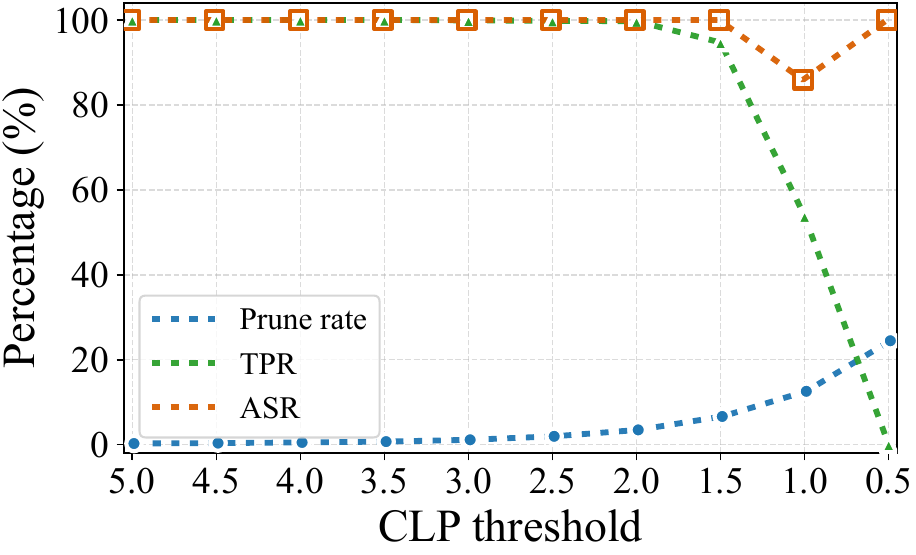}
\caption{CLP results with threshold sweeping.}
\label{fig:prc_clp}
\end{subfigure}

\caption{\mg performance under model pruning defenses.
We report pruning rate, TPR, and ASR.
The LDM is SD-2.1 and the watermarking scheme is PRCMark.
}
\label{fig:prc_prune_anp_clp}
\end{figure*}
Model pruning is a classical and widely studied defense against backdoor attacks, based on the observation that backdoor behaviors often rely on a small subset of highly influential parameters~\cite{DBLP:conf/icml/LiLMKLLJ23,DBLP:conf/raid/0017DG18}.
To evaluate the performance of pruning-based defenses against \mg, we consider three pruning methods: 
(i) channel-level structured pruning (\textbf{CLSP}), which prunes the backdoored neurons that are dormant in the presence of clean inputs~\cite{gu2017badnets},
(ii) a data-free Backdoor Removal method named \textbf{CLP}~\cite{DBLP:conf/eccv/ZhengTLL22},
(iii) an adaptive pruning defense named \textbf{ANP}\footnote{When implementing ANP~\cite{DBLP:conf/iclr/ZengCPM0J22}, we reinterpret image classifier setting into a VAE encoder setting. Concretely, we replace the classifier’s cross-entropy objective with the VAE self-reconstruction loss and use this reconstruction objective in the inner adversarial step to craft perturbations against encoder channels. More hyperparameter settings are provided in Appendix \ref{appendix: hyper-parameter}.
}~\cite{DBLP:conf/iclr/ZengCPM0J22}.
The results are shown in \Cref{fig:prc_prune_anp_clp}.
Across all three defenses, increasing pruning aggressiveness consistently reduces TPR, indicating that pruning erodes the VAE encoder needed for the benign watermark detection step by step.
In contrast, ASR stays high throughout the pruning process, with only a brief and minor fluctuation.
Overall, these trends suggest that pruning-based defenses incur noticeable functionality decrease before meaningfully suppressing the backdoor behavior, highlighting the robustness of \mg against model pruning mitigation.

\begin{figure*}[t]
\centering

\begin{subfigure}[t]{0.32\linewidth}
  \centering
  \includegraphics[width=\linewidth]{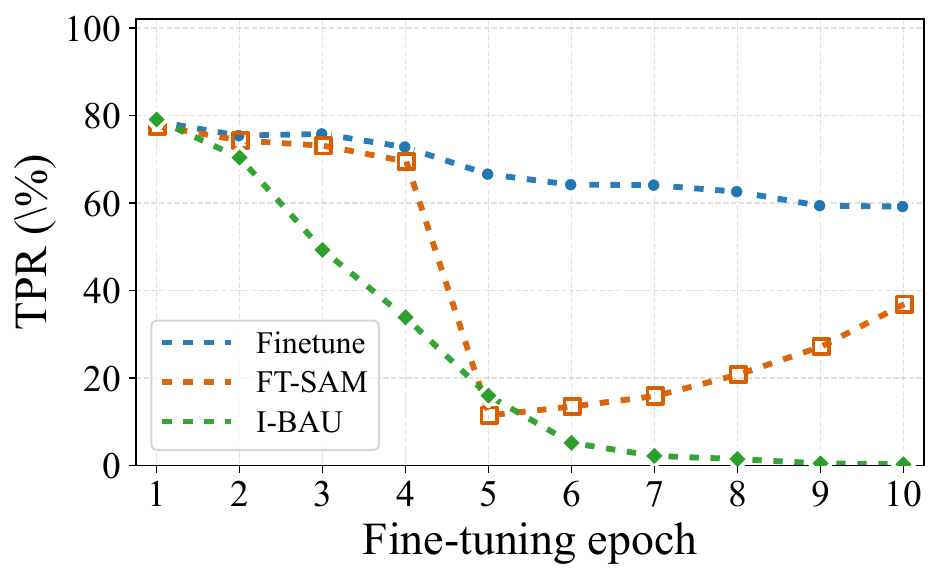}
  \caption{TPR of Tree-Ring}
\end{subfigure}\hfill
\begin{subfigure}[t]{0.32\linewidth}
  \centering
  \includegraphics[width=\linewidth]{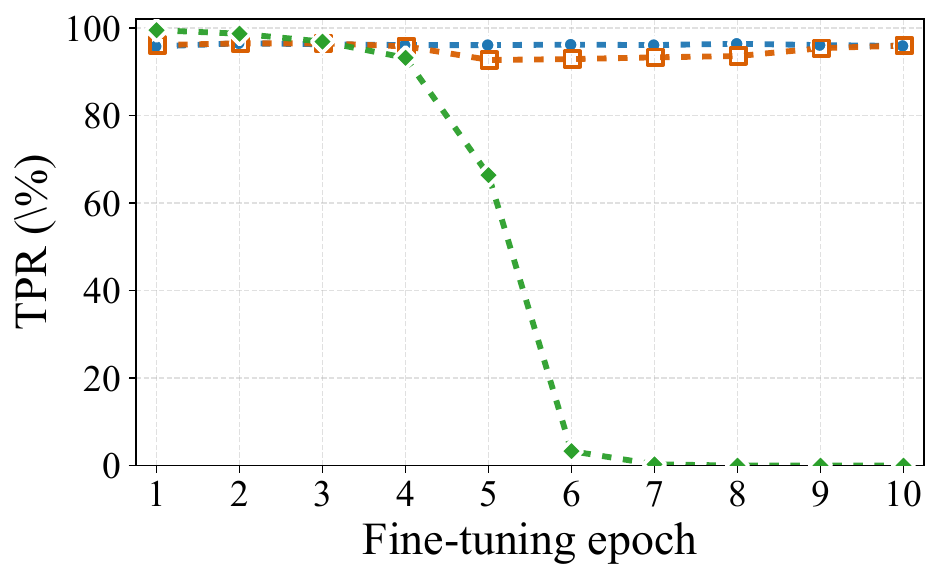}
  \caption{TPR of Gaussian Shading}
\end{subfigure}\hfill
\begin{subfigure}[t]{0.32\linewidth}
  \centering
  \includegraphics[width=\linewidth]{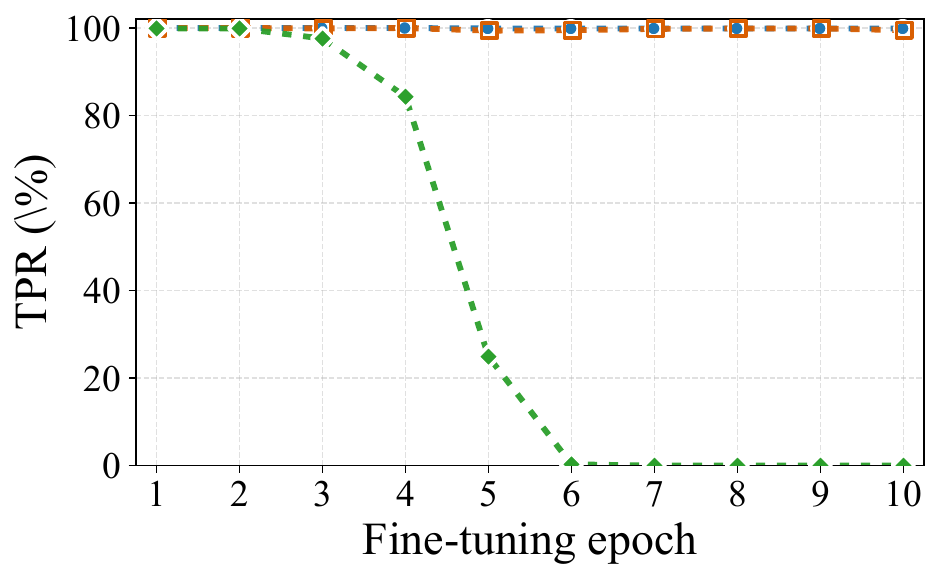}
  \caption{TPR of PRCMark}
\end{subfigure}

\begin{subfigure}[t]{0.32\linewidth}
  \centering
  \includegraphics[width=\linewidth]{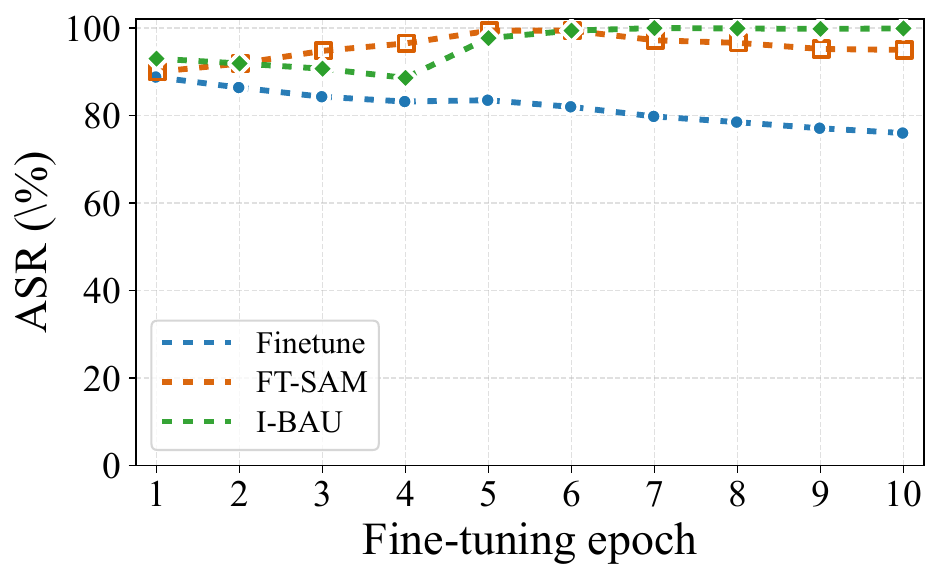}
  \caption{ASR of Tree-Ring}
\end{subfigure}\hfill
\begin{subfigure}[t]{0.32\linewidth}
  \centering
  \includegraphics[width=\linewidth]{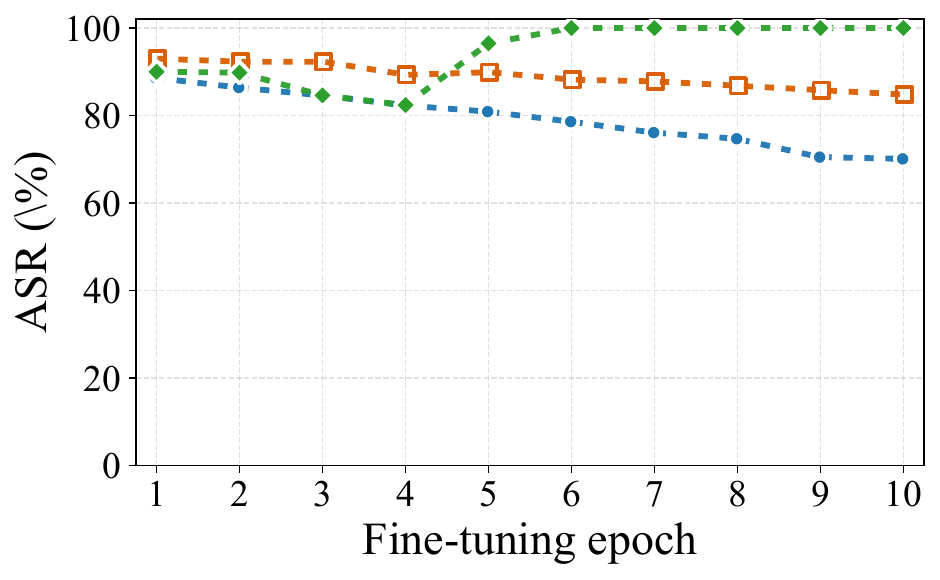}
  \caption{ASR of Gaussian Shading}
\end{subfigure}\hfill
\begin{subfigure}[t]{0.32\linewidth}
  \centering
  \includegraphics[width=\linewidth]{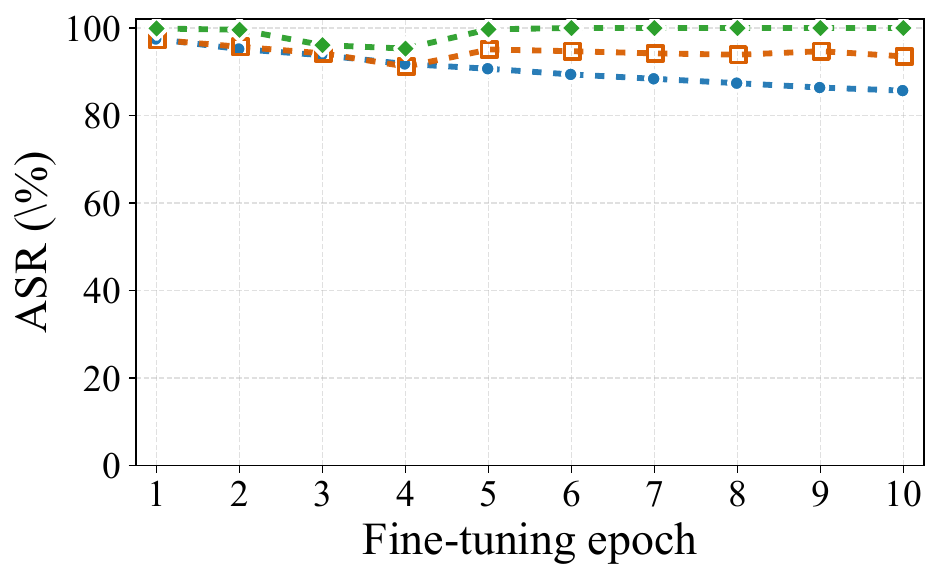}
  \caption{ASR of PRCMark}
\end{subfigure}

\caption{Fine-tuning based defense dynamics across watermark schemes.
We track TPR (top row) and ASR (bottom row) over fine-tuning epochs for FT, FT-SAM, and I-BAU.}
\label{fig:six_plots_tpr_asr}
\end{figure*}

\mypara{Model Fine-tuning}
Model fine-tuning is also a classical backdoor defense, which mitigates backdoor behaviors by retraining the model on clean data to overwrite malicious correlations~\cite{DBLP:conf/nips/MinQ0C23,DBLP:conf/iclr/XuH0Q024}.
In this part, we consider three widely used fine-tuning defenses:
(i) vanilla fine-tuning (\textbf{FT});
(ii) \textbf{FT-SAM}~\cite{DBLP:conf/iccv/ZhuW0FW23}, which augments fine-tuning with sharpness-aware minimization to suppress backdoor-relevant neurons; and
(iii) \textbf{I-BAU}~\cite{DBLP:conf/iclr/ZengCPM0J22}, an implicit backdoor adversarial unlearning method.
For all three defenses, we follow the fine-tuning objective in~\cite{Rombach2021HighResolutionIS} and fine-tune the VAE for 10 epochs on 8,000 clean images generated by SD-2.1 using the prompt sampling from the SDP dataset~\cite{gustavosta_sd_prompts}.
Additional details are provided in Appendix \ref{appendix: hyper-parameter}.
We report the TPR and ASR of \mg after implementing these three model fine-tuning defenses in \Cref{fig:six_plots_tpr_asr}.
Overall, FT weakly affects the backdoor.
For Tree-Ring, FT gradually lowers TPR and slightly reduces ASR, but the ASR remains high throughout training. 
For Gaussian Shading and PRCMark, FT largely preserves watermark detection; although ASR exhibits a mild downward trend, it remains consistently high throughout training.
FT-SAM behaves similarly on Gaussian Shading and PRCMark, where both TPR and ASR remain relatively stable with only minor fluctuations. 
For Tree-Ring, however, FT-SAM exhibits a transient effect: TPR drops sharply in the middle epochs and then partially recovers, while ASR first rises and then slightly decreases but remains high in later epochs. 
In contrast, I-BAU exhibits an unfavorable trade-off for the defender. 
As training proceeds, it substantially degrades TPR across all three watermarking schemes, while ASR remains high and increases to nearly 100\% as watermark detection itself collapses.
In summary, fine-tuning-based defenses do not reliably remove \mg: FT and FT-SAM mostly preserve the backdoor behavior, whereas I-BAU degrades normal watermark detection without removing the backdoor behavior.

\mypara{Cross-encoder Verification}
\label{sec:crossencoder}
We further consider a stronger verifier-side countermeasure that replaces the verification encoder with the same architecture but different parameters. 
Specifically, on SD-2.1, we replace \texttt{sd-vae-ft-mse}~\cite{sdvaemseft} with \texttt{sd-vae-ft-ema}~\cite{sdvaeemaft} during watermark verification. 
Under this cross-encoder setting, \mg attains ASRs of 29.5\%, 35.2\%, and 31.6\% on Tree-Ring, Gaussian Shading, and PRCMark, respectively. The corresponding TPRs remain 94.2\%, 98.9\%, and 100.0\%.
These results show that changing the verification encoder weakens the attack, but does not fully eliminate its transferability.
Moreover, we can further improve the transferability of the trigger using DI$^2$-FGSM~\cite{8953423}, which increases ASR to 61.9\%, 54.6\%, 56.5\% on Tree-Ring, Gaussian Shading, and PRCMark, respectively.
Therefore, cross-encoder verification is a meaningful mitigation, but it should not be viewed as a complete solution. 
The observed transferability indicates that the security of encoder-dependent watermark verification pipeline remains an important concern. 

\subsection{Latent-space Defenses}
\label{sec: latentdefense}

\mypara{Meta-classifier-based Detection}
Following~\cite{DBLP:conf/sp/JiaLG22}, we use Meta Neural Trojan Detection (\textbf{MNTD})~\cite{DBLP:conf/sp/XuWLBGL21} to classify backdoored and benign encoders to demonstrate the stealthiness of \mg in latent space.
To construct the training dataset of the Jumbo MNTD meta classifier, we generate 100 clean shadow encoders using the same setting as described in \Cref{sec: paradefense} and construct 100 backdoored VAEs using \mg.
On the evaluation dataset of the other 20 benign and 20 \mg encoders, the trained meta-classifier achieves 100\% accuracy on benign encoders but 0\% accuracy on \mg encoders, i.e., it consistently labels \mg as benign.
This result indicates that MNTD fails to expose the backdoor signal of \mg, further supporting its latent-space stealthiness.

\begin{figure}[t]
\centering
\includegraphics[width=0.99\linewidth]{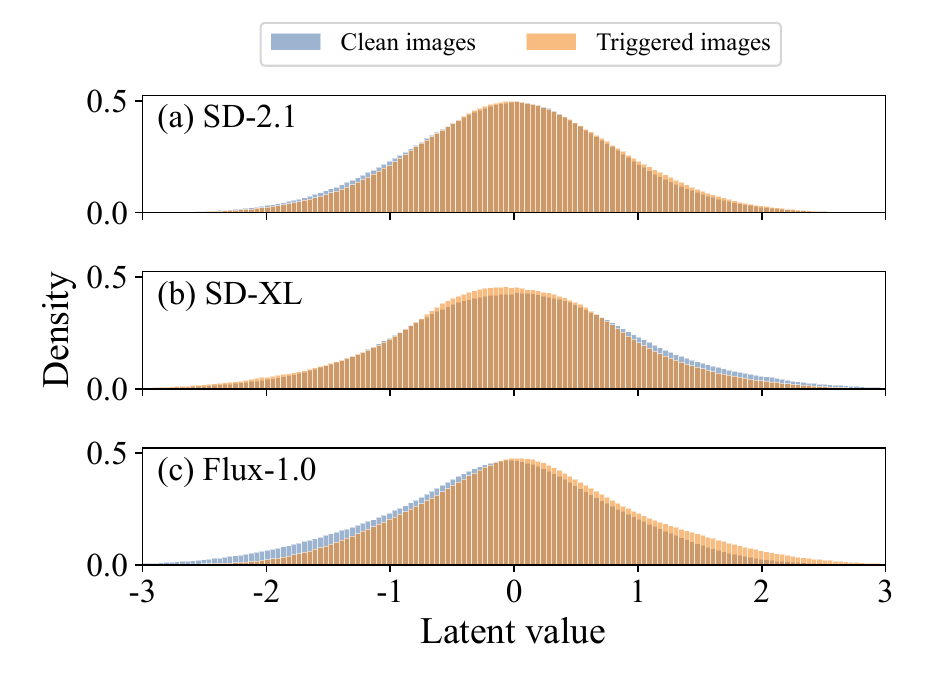}
\caption{Latent-space stealthiness of \mg.
Empirical marginal density of latent values on 1,000 PRCMark images, comparing clean inputs and triggered inputs for SD-2.1, SD-XL, and FLUX-1.0.}
\label{fig:latent_dist}
\end{figure}

\mypara{Visualization of Latent Distributions}
To further demonstrate the stealthiness of \mg in the latent space, we visualize the distribution of latents for both watermarked images and their counterparts with triggers added, shown in \Cref{fig:latent_dist}.
Owing to the sign-flipping constraint we employed, \mg preserves the global latent statistics for triggered images.
To complement this empirical visualization with a more principled statistical analysis, we examine whether latent-distribution can expose triggered inputs through a two-sample Kolmogorov-Smirnov (KS) test~\cite{hodges1958smirnov}\footnote{We use the official implementation in SciPy: \url{https://docs.scipy.org/doc/scipy/reference/generated/scipy.stats.ks_2samp.html}}. 
The first sample consists of latents from 1,000 clean PRCMark images encoded by the backdoored SD-2.1 encoder. 
The second sample consists of latents from $n$ query images, where we sweep $n$ from 1 to 100 and consider two settings in which the queries are either all clean or all triggered.
We then apply the two-sample KS test to these two samples. 
In all cases, the test rejects the null hypothesis for both clean and triggered queries. 
Therefore, this KS-based detector fails to reliably distinguish triggered inputs from clean ones, as its rejection behavior is not trigger-specific.
The above results highlight a key property of our attack: the backdoor alters functional behavior under the trigger while remaining statistically indistinguishable in latent space.

\begin{figure*}[t]
\centering
\includegraphics[width=0.99\linewidth]{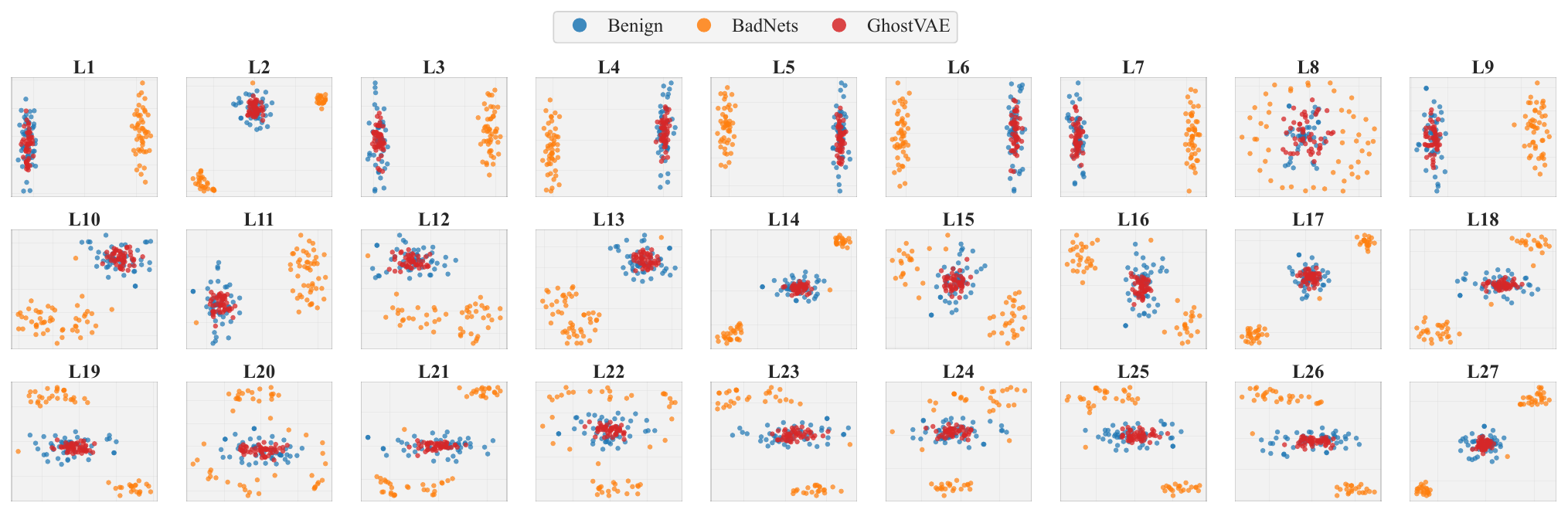}
\caption{Layer-wise t-SNE visualization of convolutional weights in the VAE encoder.
Each subplot (L1–L27) corresponds to a convolutional layer.
We fix the LDM to SD-2.1.}
\label{fig:tsne}
\end{figure*}

\mypara{Activation Clustering~\cite{DBLP:conf/aaai/ChenCBLELMS19}}
We further assess whether activation clustering can separate clean and triggered inputs to \mg. 
For this evaluation, we use 500 clean watermarked images and their corresponding triggered versions.
We extract the final encoder activations of these 1,000 inputs, reduce them to lower-dimensional activations using Independent Component Analysis (ICA)~\cite{HYVARINEN2000411}, and then apply $K$-means clustering.
We quantify separability using the mean optimal clustering accuracy over five runs.
Specifically, because the two cluster indices produced by $K$-means are arbitrary, we assign them to the ground-truth clean and triggered labels in the way that yields the highest classification accuracy for each run, and report the average accuracy across five independent runs.
On the SD-2.1 setting, the average optimal clustering accuracy is 52.6\% and 50.1\% when the ICA dimension is 256 and 512, respectively, which means that the two clusters contain nearly balanced clean and triggered samples.
We further observe the same phenomenon on SD-XL, where the corresponding accuracy is 50.0\% and 50.1\% under the same ICA dimensions.
These results indicate that activation clustering fails to reliably separate triggered inputs from clean ones.

\subsection{Parameter-Space Visualization}
\label{sec:56para_visual}
To further probe whether \mg leaves a detectable artifact footprint in the parameter space, we construct three groups of encoders, each containing 50 independently trained models: (i) benign VAE encoders obtained by fine-tuning SD-2.1 using the default training hyperparameters in \Cref{sec: paradefense}, (ii) backdoored VAE encoders using random BadNets-style trigger~\cite{gu2017badnets} as described
in \Cref{sec: latentdefense}, and (iii) backdoored encoders trained by \mg.
For each convolutional layer, we collect the corresponding weights from all encoders across three groups and jointly project them into a two-dimensional space using t-SNE~\cite{DBLP:journals/jstatsoft/PolicarSZ24}.
\Cref{fig:tsne} illustrates the resulting layer-wise visualizations, where each point corresponds to one encoder at a specific layer.
Across all layers, \mg-based backdoored encoders consistently overlap with the benign cluster and remain well separated from BadNets-based backdoored encoders.
These results indicate that \mg does not induce a layer-specific geometric signature in the weight space, which means the introduction of the MMD-based regularization in \mg effectively enhances the parameter-space stealthiness.

\section{Discussion \& Limitations}
\label{sec:discussian_and_limitations}

\subsection{Regulation Suggestions}
\label{sec:dis_regulation}
Our findings indicate that once the neural network component becomes unreliable, the effectiveness of watermarking schemes is fundamentally compromised.
Consequently, we argue that regulatory efforts should place greater emphasis on the integrity of the entire model deployment pipeline underlying watermarking schemes.
For example, VAEs could be required to be trained by trusted third parties, or their training processes could be subject to end-to-end regulatory oversight.
Meanwhile, to prevent covert replacements, cryptographic mechanisms such as hash functions~\cite{rfc1321} should be employed to ensure the integrity of deployed VAE components.

Moreover, integrity checking alone may be insufficient if adversarial perturbations can transfer across compatible encoders. 
As suggested by our cross-encoder verification results in \Cref{sec:crossencoder}, replacing the verification encoder weakens but does not fully eliminate the attack. 
Therefore, VAE training should also consider adversarial training to improve robustness to input perturbations, while deployment pipelines should reduce the transferability of such perturbations across independently trained encoders.

We further note recent progress in text-to-image diffusion models, where image generation can be achieved without the involvement of VAEs.
For instance, pMF~\cite{lu2026onesteplatentfreeimagegeneration} enables single-step image generation directly in the raw pixel space, without relying on a pretrained VAE or performing multi-step iterative sampling.
While such architectures may be inherently resilient to our attacks, watermarking schemes specifically designed for these novel generation paradigms remain largely underexplored, thereby posing challenges even for basic provenance tracking of generated images.

\subsection{Limitations}
\label{sec:limitations}
We highlight several limitations of our work. 
First, implanting the backdoor into the VAE encoder can affect benign watermark detection in some settings. 
Although \mg largely preserves the original detection capability, we observe TPR degradation for certain backbone-watermark combinations, as shown in \Cref{tab:attack_results_tpr_fpr_main}. 
This suggests a trade-off between attack effectiveness and function maintenance.
An important next step is to design backdoor attacks that further minimize such degradation.
Second, our threat model assumes that the malicious model supplier can control the encoder used during watermark detection.
If the verifier instead uses an encoder that has the same architecture but different parameters, the attack can be weakened.
However, with techniques that enhance adversarial transferability, such as DI$^2$-FGSM~\cite{8953423}, \mg can still retain a non-trivial watermark evasion ASR under cross-encoder verification. 
Further studying attacks in this stronger setting, and understanding how far such transferability can be improved, would be a valuable direction for future work.
Third, part of our evidence for the stealthiness relies on empirical visualization. 
In particular, the t-SNE visualization provides only qualitative evidence and should not be interpreted as a principled detector. 
A longer-term goal is therefore to develop watermark evasion backdoors with cryptographic undetectability guarantees~\cite{Goldwasser2022PlantingUB}.

% ----------------------------------------------------

% ----------------------------------------------------
\section{Conclusion}
% ----------------------------------------------------
We propose \mg, a stealthy VAE encoder backdoor that achieves watermark evasion under trigger activation while preserving detection behavior on benign inputs.
Moreover, the universal trigger remains effective under common corruptions, purification techniques and social media pipeline, making the evasion reliable in the real world.
Crucially, we show that \mg is difficult to detect or mitigate using existing defenses, including a wide range of parameter-space and latent-space defense strategies.
These findings underscore a fundamental gap between watermark robustness and the security of the underlying generative model, where a watermark can be robust in isolation but still fail in practice when the detection pipeline is compromised.
Therefore, safeguarding semantic watermarks requires not only robust designs but also principled protection of neural network components against backdoor insertion.
We hope this work motivates future research and regulatory efforts to jointly address watermark robustness and the secure deployment of generative models.

\section*{Acknowledgement}
We sincerely thank the anonymous reviewers for their constructive suggestions. 
This work is supported by the National Cyber Security-National Science and Technology Major Project (2026ZD1500700), the Scientific Research Innovation Capability Support Project for Young Faculty (ZYGXQNJSKYCXNLZCXM-P4), the Fundamental and Interdisciplinary Disciplines Breakthrough Plan of the Ministry of Education of China (JYB2025XDXM114), the National Natural Science Foundation of China (62402273), the Guangdong Basic and Applied Basic Research Foundation (2026A1515030046), and the State Key Laboratory of Internet Architecture, Tsinghua University (HLW2025ZD14).

%-------------------------------------------------------------------------------
\appendix

\section*{Ethical Considerations}
\label{sec:ethics}

Our ethical considerations consist of \textbf{stakeholder analysis}, \textbf{impact analysis}, \textbf{mitigations}, and \textbf{decision justifications}.

\mypara{Stakeholder Analysis}
We organize the stakeholders affected by \mg according to their role in the watermark deployment and enforcement pipeline.
(1) \emph{Governance stakeholders:} This group includes regulators, auditors, platform operators, and AI image generation companies. These stakeholders care about whether deployed LDMs satisfy watermarking-related compliance requirements.
(2) \emph{Model suppliers:} They provide LDMs to generate images. Upon receiving requests from users, these models generate watermarked images, where the watermark is only detectable by model suppliers holding the watermark key.
(3) \emph{Downstream participants:} This group includes casual end users and professional content creators who use LDMs to generate images that contain watermarks.
(4) \emph{Research team:} This group includes researchers and watermark developers who study the reliability of watermarking schemes.

\mypara{Impact Analysis}
Our analysis identifies both positive and negative impacts on the stakeholder groups above.

\emph{\textbf{Positive impacts:}}
(1) \emph{Enhancing Governance Auditing (for governance stakeholders):} Our work reveals a hidden risk of surface-level compliance, where a system appears compliant during auditing but can still have vulnerabilities in real deployment. Our work will help make future auditing processes of LDM watermarking systems more comprehensive.
(2) \emph{Strengthening Secure Deployment Practices (for model suppliers):} Our work helps model supplier recognize that watermark embedding alone is insufficient. They need to protect the entire detection pipeline, especially neural components like the VAE encoder, reducing the risk of hidden backdoors and improving trustworthy deployment.
(3) \emph{More Trustworthy Provenance (for downstream participants):} By exposing this hidden vulnerability, our work can help drive the development of more trustworthy provenance mechanisms, thereby reducing the risk that downstream participants rely on provenance signals that fail in practice.
(4) \emph{Advancing Research on Watermark Security (for research team):} Our work exposes a previously overlooked attack surface in semantic watermarking systems, motivating future research on verifier integrity, backdoor-resistant watermarking, and end-to-end security for LDM deployment pipelines.

\emph{\textbf{Negative impacts:}}
(1) \emph{Reduced Confidence in Watermark-Based Enforcement (for governance stakeholders):} Our work may weaken confidence in current watermark-based enforcement by showing that these systems can be less reliable than expected.
(2) \emph{Increased Misuse and Commercial Risks (for model suppliers):} Our work may help malicious suppliers selectively disable watermark detection, while also creating commercial risks for benign providers, such as customer churn.
(3) \emph{Harder-to-Trace Harmful Content (for downstream participants):} Our work may enable malicious users to generate images that evade watermark detection, making harmful or misleading content harder to trace.
(4) \emph{Challenging Existing Research Assumptions (for research team):} Our work may challenge reliability assumptions in existing semantic watermarking schemes by showing that watermarked latents alone are insufficient for security evaluation. While this may prompt researchers to revisit prior protocols and claims, our findings do not invalidate the underlying cryptographic assumptions; instead, they expose risks arising from the integration of cryptographic primitives with AI components.

\mypara{Mitigations}
We acknowledge that this work may be misused by malicious actors to better understand or reproduce watermark evasion attacks. For example, a malicious supplier–user coalition could use a backdoored watermarking pipeline to generate and disseminate harmful synthetic images while evading traceability. To mitigate such risks, we adopt the following measures.

As discussed in \Cref{sec:dis_regulation}, our mitigation suggestions include trusted third-party VAE training, end-to-end oversight of training and deployment, integrity checks for deployed VAE components, and designing more secure watermarking schemes. We emphasize that any use of our findings should remain within applicable law and regulation.

Meanwhile, to mitigate real-world threats, we did not release any attack-ready artifacts generated in this work, such as optimized triggers, backdoored VAE encoders, or directly reusable checkpoints. Our evaluation was restricted to controlled offline experiments on local systems, and we did not test on deployed large-scale services, engage with real users, or attempt any real-world watermark evasion or misuse. These restrictions were intended to limit operational abuse while preserving the value of the work as a security assessment.

\mypara{Decision Justifications}

\emph{\textbf{Decision to Conduct:}}
It was appropriate to conduct this work because the vulnerability we studied is a practical deployment-level risk in semantic watermarking systems whenever detection relies on neural components that can be backdoored. Our motivation was to assess whether this attack is practically feasible and whether it can remain stealthy in deployment. Before starting this work, we explicitly considered the relevant stakeholders, the misuse risks, and the conditions under which the work could be responsibly conducted. We also carried out all experiments in a controlled offline setting.

\emph{\textbf{Decision to Publish:}}
Although backdoor research on general AI models has been extensively published, its integration with watermarking scenarios has not yet been well studied. Without disclosure, this vulnerability could remain hidden and sustain misplaced confidence in watermark-based enforcement. After weighing the misuse risks against the defensive value of disclosure, we concluded that publication is justified, as it can help the community avoid false assurance, motivate more secure watermarking schemes, and strengthen the auditing process. We believe the benefits of publication outweigh the risks, and by sharing our findings through a high-impact platform like USENIX, we can promote long-term security of the watermarking ecosystem.

\section*{Open Science}
To support reproducibility, we release an artifact package that contains (i) the necessary configuration files, (ii) all defense and evaluation scripts used in our study, and (iii) the full implementation of \mg.
Specifically, the repository provides an integrated workflow, including environment setup, dataset preparation, universal-trigger optimization, backdoored VAE-encoder training, and representative defense runs.

In particular, we provide two standard dependency specifications, \texttt{requirements\_dif.txt} and \texttt{requirements\_ghostvae.txt}, for different task stages, together with example commands.
The main results in this paper can be reproduced by following the instructions in \texttt{README.md}.
All artifacts are available via \url{https://github.com/CryptoAILab/GhostVAE} and \url{https://doi.org/10.5281/zenodo.20391190}.

\bibliographystyle{plainurl}
\bibliography{ref}

\appendix

\section{RAPSD Calculation}
\label{appendix: rapsd}

Given an image $x$ and its triggered counterpart $x+\mathcal{G}(\delta;\sigma)$, we summarize their frequency content using the radially-averaged power spectral density (RAPSD).
For an input image $x$, we compute its Fourier frequency matrix
$f := F(x)\in\mathbb{C}^{h\times w}$,
where $h$ and $w$ denote the height and width of the Fourier grid (matching the spatial resolution of $x$), and $(u,v)$ indexes a discrete frequency coordinate on this $h\times w$ grid.
We then partition the coordinates of $f$ into $B$ radial bands according to their distance to the spectrum center, yielding $B$ disjoint sets
$\{\mathcal{B}_b\}_{b=1}^{B}$, where $\mathcal{B}_b$ contains the frequency coordinates that fall into the $b$-th band and $|\mathcal{B}_b|$ denotes its cardinality.
RAPSD is defined as a $B$-dimensional vector whose $b$-th entry is the average squared magnitude of FFT coefficients within $\mathcal{B}_b$:
\begin{equation}
\mathrm{RAPSD}_b(f)
\;:=\;
\frac{1}{|\mathcal{B}_b|}
\sum_{(u,v)\in\mathcal{B}_b} |f(u,v)|^2.
\label{eq:rapsd_def}
\end{equation}
Collecting all $B$ band-wise entries yields the RAPSD vector.

\section{Experimental Details}

\subsection{Watermark Detection Threshold Setting}
\label{appendix: threshold-setting}
We summarize the detection thresholds for each semantic watermarking on each backbone model.
These thresholds shown in \Cref{tab:wm_thresholds} are fixed across all experiments in our evaluation.
Different models and watermark designs induce different calibration characteristics; thus, a single global threshold would conflate detection stringency across settings and lead to unfair comparisons.
We use the official thresholds when they are specified in the original watermarking papers; for other backbone-watermark combinations, we calibrate thresholds to achieve high TPR and low FPR on benign models. We further assess threshold sensitivity on Gaussian Shading with SD-XL by varying the threshold from 0.67 to 0.60 and 0.70. TPR/FPR/ASR change to 73.0\%/0.0\%/100.0\% and 99.6\%/1.3\%/92.5\%, respectively. While these changes reflect the expected threshold trade-off, \mg consistently maintains strong evasion performance, indicating that our findings are robust to reasonable threshold choices.

\begin{table}[t]
\centering
\caption{Detection threshold configurations for watermark verification.
For PRCMark, we show the $\sigma_{prc}$ here.}
\label{tab:wm_thresholds}
\small
\setlength{\tabcolsep}{6pt}
\renewcommand{\arraystretch}{1.15}
\begin{tabular}{lccc}
\toprule
\multirow{2}{*}{Scheme} & \multicolumn{3}{c}{Threshold} \\
\cmidrule(lr){2-4}
& SD-2.1 & SD-XL & Flux-1.0 \\
\midrule
Tree-Ring       & 0.59 & 0.67 & 0.60 \\
Gaussian Shading  & 1.00 & 0.98 & 0.95 \\
PRCMark          & 0.15 & 0.15 & 0.15 \\
\bottomrule
\end{tabular}
\end{table}
\begin{table}[t]
\centering
\caption{Hyperparameter settings for parameter-level backdoor defenses.}
\label{tab:defense_hparams_compact}
\small
\setlength{\tabcolsep}{4.5pt}
\begin{tabular}{p{0.18\linewidth} p{0.74\linewidth}}
\toprule
Defense & Hyperparameters  \\
\midrule
CLSP &   pruning rate = $0.15\%$, prune steps = $20$ \\
\midrule
CLP &    sweep pruning threshold from $0.5$ to $5.0$ with step size $0.5$ \\
\midrule
ANP &
epochs=4, batch size=2, $\epsilon$ = 0.4, $lr_{inner}$ = 0.4, $lr_{outer}$ = 0.2, sweep pruning threshold from $0.1$ to $0.9$ with step size $0.1$\\
\midrule
FT & epoch = 10, $lr=1e-5$ \\
\midrule
FT-SAM &  epoch = 10, $lr=1e-5$, $\rho_{\max}=2.0$, $\rho_{\min}=0.05$, $\rho_{schedule} = cosine$\\
\midrule
I-BAU & epoch = 10, $lr=1e-5$, $n_{rounds}$ = 4, K = 5, pert steps = 50, unlearn portion = 0.01, unlearn batches = 300\\
\midrule
KS test & significance level $\alpha=0.05$\\
\bottomrule
\end{tabular}
\end{table}

\subsection{Defense Methods Parameter Setting}
\label{appendix: hyper-parameter}
We disclose the hyperparameter settings for the model pruning and fine-tuning defenses evaluated in \Cref{sec: paradefense}.
\Cref{tab:defense_hparams_compact} lists the exact configurations used in our experiments.
Compared to the default hyperparameter configuration, we reduce the number of epochs because training the VAE encoder is computationally time-consuming.

\begin{figure}[t]
\centering
\includegraphics[width=0.7\linewidth]{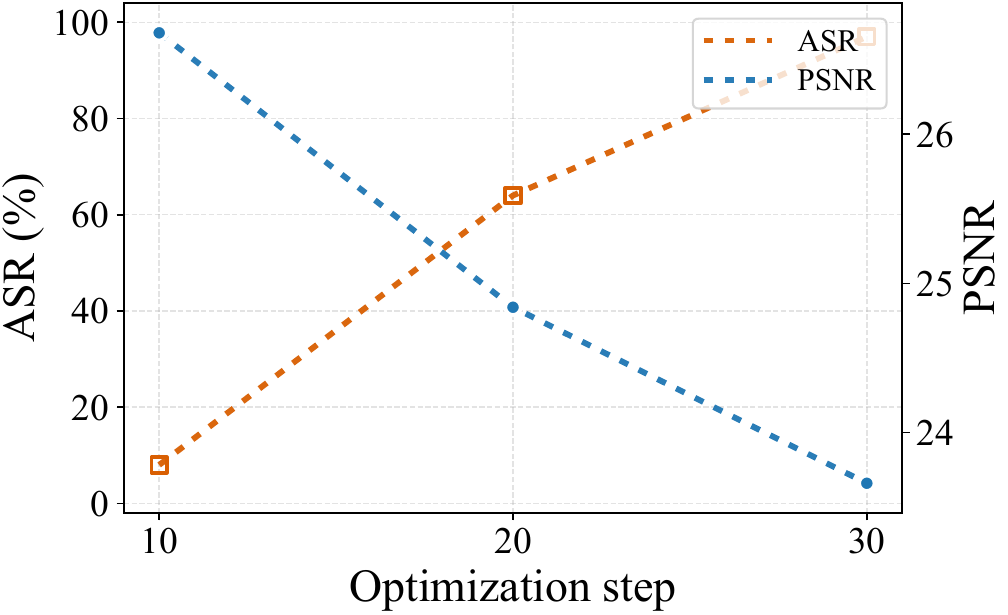}
\caption{Sensitivity of the Imprint attack to the optimization steps $T_{\text{imprint}}$.
We report the resulting ASR (left axis) and image quality measured by PSNR (right axis) as $T_{\text{imprint}}$ varies.}
\label{fig: removal_imp}
\end{figure}

\begin{figure}[t]
\centering
\includegraphics[width=0.7\linewidth]{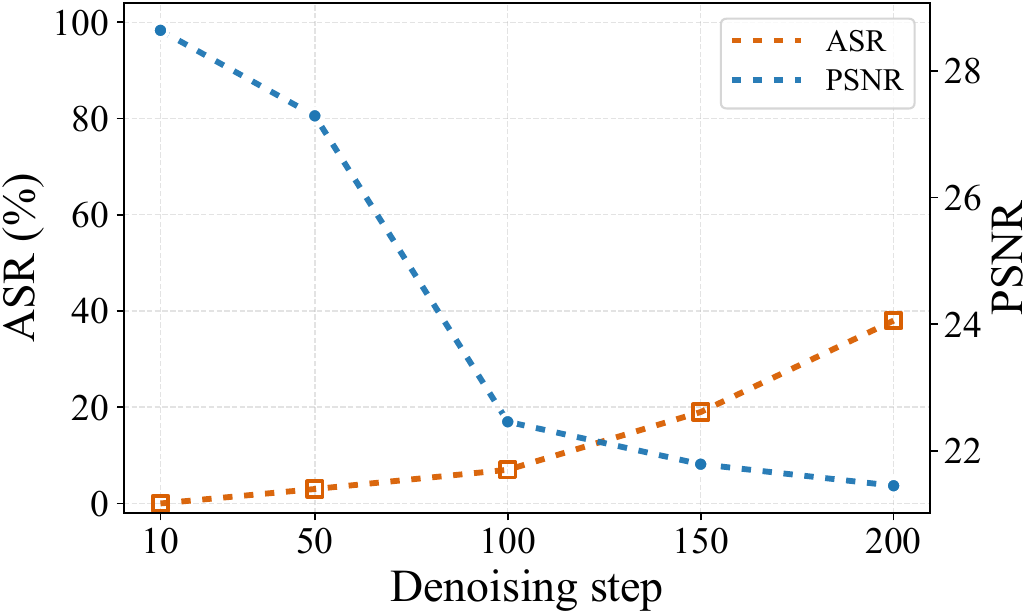}
\caption{Sensitivity of the Regeneration attack to the denoising steps $T_{\text{reg}}$.
We report the resulting ASR (left axis) and image quality measured by PSNR (right axis) as $T_{\text{reg}}$ varies.}
\label{fig: removal_reg}
\end{figure}

\section{Sensitivity to Step Budgets for Watermark Removal Attacks}
\label{appendix:removal_steps}

We perform a sensitivity study on the step budgets of two iterative watermark removal baselines: the optimization steps $T_{\text{imprint}}$ used by Imprint and the denoising steps $T_{\text{reg}}$ used by Regeneration.
We sweep each budget while holding all other settings fixed, and report the ASR and image quality.

Both iterative baselines exhibit a clear trade-off: increasing the step budget improves watermark evasion but degrades image quality and increases runtime.
For Imprint attack results shown in \Cref{fig: removal_imp}, raising the optimization steps from $T_{\text{imprint}}{=}10$ to $30$ sharply increases the ASR from $3\%$ to $97\%$, but this gain comes with a substantial PSNR drop, indicating noticeably stronger visual distortion.
For Regeneration, \Cref{fig: removal_reg} shows the same pattern: larger denoising budgets increase ASR monotonically, yet PSNR steadily deteriorates, so the attack becomes more destructive and costly as it becomes more effective.
These results motivate our main-paper choices ($T_{\text{imprint}}{=}30$ and $T_{\text{reg}}{=}200$), which place both baselines in their stronger regimes while making the associated quality and efficiency costs explicit.

%%%%%%%%%%%%%%%%%%%%%%%%%%%%%%%%%%%%%%%%%%%%%%%%%%%%%%%%%%%%%%%%%%%%%%%%%%%%%%%%
\end{document}